\documentclass[conference]{IEEEtran}

\renewcommand\IEEEkeywordsname{Index Terms}
\usepackage{graphicx}
\usepackage{subcaption}
\usepackage{tikz,xparse}
\usetikzlibrary{dsp,chains}
\usetikzlibrary{matrix}
\usepackage{mathptmx}
\usepackage{verbatim}
\usepackage{calc}
\usepackage{ifthen}
\usepackage{xifthen}
\usepackage{cancel}

\usepackage{bm}
\usepackage{verbatim}
\usepackage{multirow}
\usepackage{cite}

\usepackage[nolist]{acronym} 
\usepackage{pgfplots}
\usetikzlibrary{arrows, arrows.meta,shapes,graphs,graphs.standard,quotes}
\usetikzlibrary{matrix}
\pgfplotsset{compat=newest}
\usetikzlibrary{spy}
\usetikzlibrary{patterns}

\pgfplotsset{improved scaled ticks/.style 2 args={
		scaled #1 ticks=#2,
		extra #1 tick style={
			scaled #1 ticks=#2,
			#1tick scale label code/.code=,
		},
	}}
	
\usepackage[hyphens]{url}

\usepackage[bookmarks=false]{hyperref}
\usepackage{units}
\usepackage{amsmath, amsbsy, amssymb, latexsym }
\hypersetup{bookmarksdepth=-2}
\usepackage{comment}
\usepackage[utf8]{inputenc}
\usepackage{xcolor}
\usepackage{enumitem}

\usepackage{algorithm}

\usepackage{algpseudocode}

\usepackage{etoolbox} 

\usepackage{relsize}  

\newcommand\bigPi[1][3]{\mathop{\vcenter{\hbox{%
				\larger[#1]{$\displaystyle\Pi$}}}}}

\tikzset{>=latex}

\algnewcommand\algorithmicforeach{\textbf{for each}}
\algdef{S}[FOR]{ForEach}[1]{\algorithmicforeach\ #1\ \algorithmicdo}

\algnewcommand\algorithmicswitch{\textbf{switch}}
\algnewcommand\algorithmiccase{\textbf{case}}
\algnewcommand\algorithmicassert{\texttt{assert}}
\algnewcommand\Assert[1]{\State \algorithmicassert(#1)}%
\algdef{SE}[SWITCH]{Switch}{EndSwitch}[1]{\algorithmicswitch\ #1\ \algorithmicdo}{\algorithmicend\ \algorithmicswitch}%
\algdef{SE}[CASE]{Case}{EndCase}[1]{\algorithmiccase\ #1}{\algorithmicend\ \algorithmiccase}%

\algtext*{EndCase}%


\definecolor{mittelblau}{RGB}{0, 126, 198}
\definecolor{violettblau}{cmyk}{0.9, 0.6, 0, 0}
\definecolor{rot}{RGB}{238, 28 35}
\definecolor{apfelgruen}{RGB}{140, 198, 62}
\definecolor{gelb}{RGB}{255, 229, 0}
\definecolor{orange}{RGB}{244, 111, 33}
\definecolor{pink}{RGB}{237, 0, 140}
\definecolor{lila}{RGB}{128, 10, 145}
\definecolor{hellgrau}{RGB}{224, 224, 224}
\definecolor{mittelgrau}{RGB}{128, 128, 128}
\definecolor{dunkelgrau}{RGB}{80,80,80}
\definecolor{anthrazit}{RGB}{19, 31, 31}
\definecolor{darkgreen}{RGB}{34,139,34}
\definecolor{ahmedyellow}{RGB}{204,153,0}
\definecolor{ahmedpurple}{RGB}{126,47,142}
\definecolor{ahmedsky}{RGB}{77,190,238}
\definecolor{darkbrown}{RGB}{162,20,47}
\colorlet{Mycolor1}{green!10!orange!90!}
    
\tikzset{
        vnds/.style={ shape=circle, fill=black, draw, inner sep=0pt,minimum size=10pt},
        cnds/.style={ shape=rectangle, fill=white!90!magenta, draw, inner sep=0pt,minimum size=10pt}, 
        vndGs/.style={ shape=circle, fill=darkgreen, draw=black, inner sep=1pt,minimum size=10pt},
        vndGsg/.style={ shape=circle, draw=black, fill=green, draw, inner sep=0.05pt,minimum size=5pt},
        vndRs/.style={ shape=circle, fill=red, draw, inner sep=0pt,minimum size=5pt},     
        vndRsr/.style={ shape=circle, draw=red, fill=red, draw, inner sep=0pt,minimum size=5pt},     
        vndsc/.style={ shape=circle, fill=black, draw, inner sep=0pt,minimum size=7.5pt},
        cndsc/.style={ shape=rectangle, fill=white, draw, inner sep=0pt,minimum size=7.5pt}, 
        vndGsc/.style={ shape=circle, fill=green, draw, inner sep=0pt,minimum size=7.5pt},
        vndRsc/.style={ shape=circle, fill=red, draw, inner sep=0pt,minimum size=7.5pt}
}
\def\angleRot{0} \def\vndPos{left} \def\cndPos{right} \def\scale{0.18}

\IEEEoverridecommandlockouts

\begin{document}

\begin{NoHyper}
\title{Decoder-in-the-Loop: Genetic Optimization-based LDPC Code Design}

\author{\IEEEauthorblockN{Ahmed Elkelesh\IEEEauthorrefmark{1}, Moustafa Ebada\IEEEauthorrefmark{1}, Sebastian Cammerer\IEEEauthorrefmark{1}, Laurent Schmalen\IEEEauthorrefmark{2} and Stephan ten Brink\IEEEauthorrefmark{1}} \thanks{This work has been supported by DFG, Germany, under grant BR 3205/5-1.}
\IEEEauthorblockA{
	\IEEEauthorrefmark{1} Institute of Telecommunications, Pfaffenwaldring 47, University of  Stuttgart, 70569 Stuttgart, Germany 
	\\
	\IEEEauthorrefmark{2} Communications Engineering Lab, Karlsruhe Institute of Technology, Kreuzstr. 11, 76133 Karlsruhe, Germany
}
}

{}
{}
\makeatother

\maketitle

\begin{acronym}
 \acro{ECC}{error-correcting code}
 \acro{HDD}{hard decision decoding}
 \acro{SDD}{soft decision decoding}
 \acro{ML}{maximum-likelihood}
 \acro{GPU}{graphical processing unit}
 \acro{BP}{belief propagation}
 \acro{BPL}{belief propagation list}
 \acro{LDPC}{Low-Density Parity-Check}
  \acro{HDPC}{high density parity check}
 \acro{BER}{bit error rate}
 \acro{SNR}{signal-to-noise ratio}
 \acro{BPSK}{binary phase shift keying}
 \acro{AWGN}{additive white Gaussian noise}
  \acro{bi-AWGN}{binary-input additive white Gaussian noise}
 \acro{MSE}{mean squared error}
 \acro{LLR}{Log-likelihood ratio}
 \acro{MAP}{maximum a posteriori}
 \acro{NE}{normalized error}
 \acro{BLER}{block error rate}
 \acro{PE}{processing elements}
 \acro{SCL}{successive cancellation list}
 \acro{SC}{successive cancellation}
 \acro{BI-DMC}{Binary Input Discrete Memoryless Channel}
 \acro{CRC}{cyclic redundancy check}
 \acro{BEC}{Binary Erasure Channel}
 \acro{BSC}{Binary Symmetric Channel}
 \acro{BCH}{Bose-Chaudhuri-Hocquenghem}
 \acro{RM}{Reed--Muller}
 \acro{RS}{Reed-Solomon}
  \acro{SISO}{soft-in/soft-out}
\acro{PSCL}{partitioned successive cancellation list}
  \acro{3GPP}{3rd Generation Partnership Project }
  \acro{eMBB}{enhanced Mobile Broadband}
      \acro{CN}{check node}
      \acro{VN}{variable node}      
      \acro{PC}{parity-check}
      \acro{GenAlg}{Genetic Algorithm}
\acro{AI}{Artificial Intelligence}
\acro{MC}{Monte Carlo}
\acro{CSI}{Channel State Information}
\acro{PSCL}{partitioned successive cancellation list}
\acro{IRA}{irregular repeat-accumulate}
\acro{TB-IRA}{tailbiting irregular repeat-accumulate}
\acro{PEG}{Progressive Edge Growth}
 \acro{VND}{variable node decoder}
 \acro{CND}{check node decoder}
 \acro{PTB-IRA}{pseudo-tailbiting irregular repeat-accumulate}
       \acro{GenAlg}{Genetic Algorithm}
       \acro{OSD}{ordered statistic decoding}
       \acro{OFDM}{Orthogonal Frequency-Division Multiplexing}
\end{acronym}

\begin{abstract}
LDPC code design tools typically rely on asymptotic code behavior and are affected by an unavoidable performance degradation due to model imperfections in the short length regime.
We propose an LDPC code design scheme based on an evolutionary algorithm, the \ac{GenAlg}, implementing a ``decoder-in-the-loop'' concept.
It inherently takes into consideration the channel, code length and the number of iterations while optimizing the error-rate of the actual decoder hardware architecture. 
We construct short length LDPC codes (i.e., the parity-check matrix) with error-rate performance comparable to, or even outperforming that of well-designed standardized short length LDPC codes over both AWGN and Rayleigh fading channels.
Our proposed algorithm can be used to design LDPC codes with special graph structures (e.g., accumulator-based codes) to facilitate the encoding step, or to satisfy any other practical requirement.
Moreover, GenAlg can be used to design LDPC codes with the aim of reducing decoding latency and complexity, leading to coding gains of up to $\unit[0.325]{dB}$ and $\unit[0.8]{dB}$ at BLER of $10^{-5}$ for both AWGN and Rayleigh fading channels, respectively, when compared to state-of-the-art short LDPC codes.
Also, we analyze what can be learned from the resulting codes and, as such, the GenAlg particularly highlights design paradigms of short length LDPC codes (e.g., codes with degree-1 variable nodes obtain very good results).

\end{abstract}

\vspace{0.3cm}


\acresetall

\section{Introduction}

The design of \ac{LDPC} codes is well-established at the limits of the infinite length regime. The classical design tools, e.g., density evolution \cite{densityEvol} and EXIT charts \cite{LDPCEXITStB}, provide the required analysis to design long LDPC codes of superior performance at negligible gaps from the Shannon limit \cite{CapLDPC}. However, for finite length LDPC codes, a sufficiently \emph{good} LDPC code turns out to deviate from the guidelines (e.g., degree distributions) provided by the classical design tools which are based on the asymptotic length code analysis. 
It turns out that \ac{LDPC} codes lack performance in the ultra-short length regime when compared to more structured, and thus \emph{explicit}, coding schemes such as Polar, \ac{RM} or \ac{BCH} codes (see \cite{liva2016} for an exhaustive comparison).

Nonetheless, \ac{LDPC} codes can be seen as the workhorse of many of today's (and upcoming) communication standards motivated by a simple and well-understood decoder, namely the \ac{BP} decoder. However, emerging applications based on short block transmission have urged the need for well-designed ``ultra-short'' codes, cf. ultra-reliable and low-latency communications (URLLC); e.g., for machine-to-machine type communications and Internet of Things networks. In these applications, it is also preferable to work with a unified decoding hardware, i.e., one (de-)coding scheme \emph{fits all} -- from block lengths of several hundred up to ten-thousands of bits.
This trend is also reflected by the fact that the 3GPP group agreed to replace the Turbo codes by \ac{LDPC} codes in the upcoming New Radio (NR) access technology standard \cite{5GLDPC,Richardson5G}. Also, short LDPC codes are used in near-earth and deep space applications \cite{CCSDSapplication}.
Thus, rather than finding new coding schemes for URLLC implying new decoding algorithms and hardware structures, we aim to leverage \emph{short-length} \ac{LDPC} codes by explicitly optimizing them at short length and under actual decoder constraints.
To some extent, our approach also follows the current trend of \emph{data-driven} computation/optimization in the field of machine learning, i.e., rather than tailoring the code to a specific canonical model (e.g., \ac{AWGN} channel with non-quantized messages), our method inherently takes any practical decoder hardware constraints ``in-the-loop'' into account and directly optimizes from the data (i.e., the \emph{actual} decoder behavior).

Typically, the LDPC code design is divided into two sub-problems: 1.) finding a good general code structure (degree profile or protograph) and 2.) optimizing the explicit realization of the code (edges of the actual graph). Although there exists some work on short-length code design, e.g., \cite{sEXIT}, most practical approaches rely on heuristics, e.g., greedy-based optimization techniques such as \ac{PEG} \cite{PEG} which, however, typically require degree profiles found by asymptotic assumptions. In \cite{diffEvol}, a differential evolution algorithm-based approach has been used to optimize the protograph of an \ac{LDPC} code. However, to the best of our knowledge, no differential evolution-based optimization of the \emph{full} $\mathbf{H}$-matrix has been reported so far, probably, due to the demanding computational complexity of the algorithm.  
In \cite{OSD1}, it has been shown that a simple concatenation of an \ac{LDPC} code with a \ac{CRC} code significantly enhances its performance under high-complexity \ac{OSD}. Unfortunately, the gain vanishes for \emph{classical} iterative decoding. 
Thus, it seems as if finding sparse graphs with good short length performance remains to be a cumbersome task. Yet, there is simply no suitable design strategy to find such a sparse $\mathbf{H}$-matrix due to the exponentially increasing design-space of the problem. 

A practical coding scheme also implies some further constraints on the parity-check matrix to enable low-complexity encoding such as, e.g., accumulator-based structures. We show that our approach can also be applied to given code structures such as \ac{IRA} codes. We refer to \cite{ LivaStructuredLDPC} for details on these structural graph constraints.

The main contribution of this work is an efficient LDPC code design tool resulting in short codes comparable to, or even outperforming, state-of-the-art short LDPC codes over both AWGN and Rayleigh fading channels. 
The proposed scheme is used to design the complete parity-check matrix (i.e., $\mathbf{H}$-matrix) directly, unlike the classical way where the degree distribution is optimized first. 
This optimization involves no \ac{PEG} \cite{PEG} or similar algorithms, but is only based on the \ac{GenAlg}, similar to what has been proposed for polar codes in \cite{GenAlg_Journal}.
One strong asset of the proposed design tool is that it can be tailored to any specific required constraint on the $\mathbf{H}$-matrix, resulting in codes which are of both low encoding and, thus, low hardware complexity. 
It is worth mentioning that extensions to longer LDPC codes are straightforward with the current framework. Furthermore, designing LDPC codes tailored to other types of decoders (e.g., quantized BP decoder, OSD) is possible. The source code and the $\mathbf{H}$-matrices from this work can be found online.\footnote{\url{https://github.com/AhmedElkelesh/Genetic-Algorithm-based-LDPC-Code-Design}}

\section{LDPC Codes}

An LDPC code, originally introduced by Gallager \cite{Gallager}, is conventionally represented by its corresponding $(m\times n)$ parity-check matrix $\mathbf{H}=\left[h_{ji}\right]_{m\times n}$ (referred to as $\mathbf{H}$-matrix throughout this work), where $n$ represents the number of \acp{VN} (i.e., also the code block length) and $m$ represents the number of \acp{CN} the code has. The number of information bits per codeword is $k=n-\text{rank}\left(\mathbf{H}\right)$. Therefore, the actual code rate is designated by $R_c=\nicefrac{k}{n}$ which could be potentially higher than the so-called design rate $r_d=\nicefrac{(n-m)}{n}$. A corresponding graphical representation \cite{Tanner} is the Tanner graph, in which a VN $v_i$ is connected to a CN $c_j$ if $h_{ji}=1$, with $i\in\{1,\dots,n\}$ and $j\in\{1,\dots,m\}$.

The decoding of LDPC codes is iteratively performed over the Tanner graph where soft messages (i.e., Log Likelihood Ratio (LLR) messages) are propagated over the graph between variable nodes and check nodes according to 

\vspace{-0.15cm}

\begin{equation*}\label{eqLcv}
	L_{c_j\rightarrow v_i}=2\cdot\tanh^{-1} \left(\bigPi_{i'\neq i}\tanh \left( \frac{L_{v_{i'}\rightarrow c_j}}{2} \right) \right)
\end{equation*}
\begin{equation*}\label{eqLvc}
L_{v_i\rightarrow c_j}=L_{ch,i}+\displaystyle\sum_{j'\neq j}L_{c_{j'}\rightarrow v_i}
\end{equation*}
where $L_{ch,i}$ is the LLR channel output, $L_{c_j\rightarrow v_i}$ is the message from CN $c_j$ to VN $v_i$ and $L_{v_i\rightarrow c_j}$ is the message from VN $v_i$ to CN $c_j$.
For more details, we refer to \cite{LDPC_maxLog,LivaStructuredLDPC} and \cite{shuLin}.

LDPC code design is, thus, the process of determining (i.e., optimizing) the connections $h_{ji}\in\{0,1\}$ of the bipartite Tanner graph under certain requirements (e.g., a target error floor or some hardware constraints). Optimizing the degree distributions of the Tanner graph is conventionally pursued via EXIT charts \cite{LDPCEXITStB}, by matching the EXIT curves of the \ac{CND} and \ac{VND}. This means that the open decoding tunnel between the two EXIT curves should be minimal to operate close to the channel capacity \cite{LDPCEXITStB}. Another method, density evolution \cite{densityEvol}, iteratively tracks the (average) probability density functions of the messages propagated between the VND and CND.

\def\scale{0.48}
\begin{figure}[t]
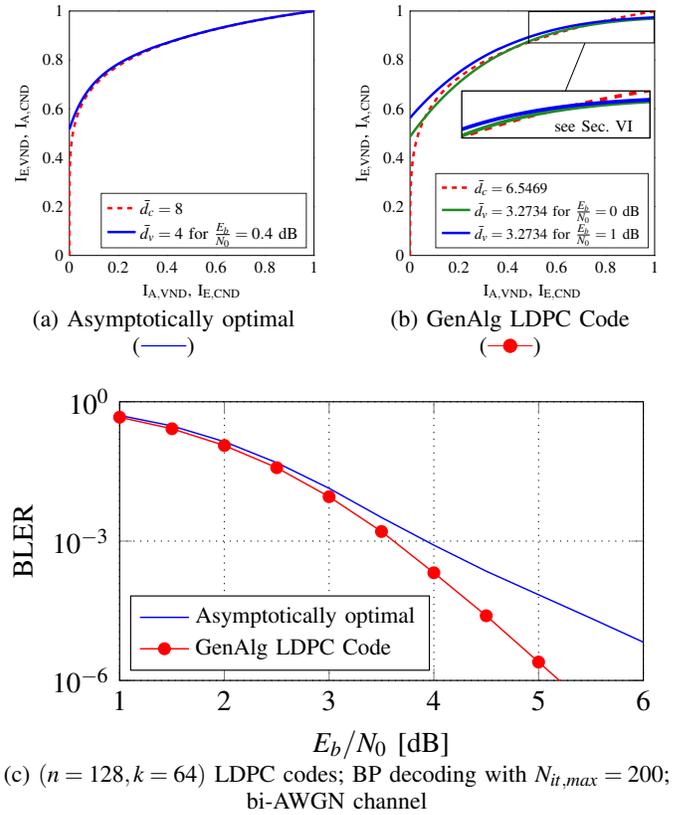

	\centering
	\captionsetup[subfigure]{justification=centering}
	\begin{subfigure}{\scale\columnwidth}
		\resizebox{\columnwidth}{!}{\input{./tikz/Matched_EXIT_17Apr2019.tikz}} 
\vspace{-0.6cm}
		\caption{Asymptotically optimal \\ 
			(\ref{lab:BLER_matched})} 
		\label{fig:EXIT_matched}	
	\end{subfigure}            \vspace{0.25cm}
	\hfill
	\begin{subfigure}{\scale\columnwidth}
		\resizebox{\columnwidth}{!}{\input{./tikz/EXIT_Random_H_200_Iter_GenAlg_des5dB.tikz}} 
\vspace{-0.6cm}		
		\caption{GenAlg LDPC Code \\ (\ref{plot:GenAlg_200_5dB})} \label{fig:GenAlg_EXIT}
	\end{subfigure}
	\begin{subfigure}{\columnwidth}
		\resizebox{\columnwidth}{!}{\begin{tikzpicture}
\begin{semilogyaxis}[
     width=8cm,
     height=5cm,
     grid=both,
          grid style={dotted,anthrazit},
       legend cell align=left,
       legend style={font=\footnotesize},
       legend columns=1,
       legend style={at={(0.02,0.17)},anchor=west,draw,
       	/tikz/every even column/.append style={column sep=0.5cm}
       },
	xmin=1,
	xmax=6,
	ymin=1e-6,
	ymax=1,
	xlabel={$E_b/N_0$ [dB]},
	ylabel={BLER},
	legend image post style={mark indices={}},
	mark options={solid}
	]

\addplot[color=blue,line width = 0.5pt,solid]
table[row sep=crcr]{
1	0.508410000000000\\
1.5	0.301550000000000\\
2	0.138250000000000\\
2.5	0.0487200000000000\\
3	0.0136800000000000\\
3.5	0.00322760000000000\\
4	0.000819000000000000\\
4.5	0.000224933333333333\\
5	6.99400000000000e-05\\
5.5	2.18600000000000e-05\\
6	6.68000000000000e-06\\
};
\label{lab:BLER_matched}
\addlegendentry{Asymptotically optimal}; 

\addplot [color=red,line width = 0.5pt,solid,mark=*]
table[row sep=crcr]{
	1	0.464370000000000\\
	1.5	0.262120000000000\\
	2	0.114480000000000\\
	2.5	0.0380000000000000\\
	3	0.00903000000000000\\
	3.5	0.00160840000000000\\
	4	0.000208000000000000\\
	4.5	2.47000000000000e-05\\
	5	2.49000000000000e-06\\
	5.5	2.50000000000000e-07\\
};
\addlegendentry{GenAlg LDPC Code};

\end{semilogyaxis}
\end{tikzpicture}} 
		\vspace{-0.7cm}
		\caption{$\left(n=128,k=64\right)$ LDPC codes; BP decoding with $N_{it,max}=200$; bi-AWGN channel} \label{fig:BLER_36_GenAlg}	
	\end{subfigure}
	\vspace{-0.1cm}
	\caption{\footnotesize The LDPC code design paradigm for asymptotic length fails in the short-length regime.}
	\vspace{-0.8cm}
	\label{fig:shortlength}
\end{figure}

The classical design methods assume infinitely long lengths $n\to\infty$, a graph that contains no cycles and infinite number of decoding iterations \cite{finiteIter}. These assumptions are not valid when considering the problem of short \ac{LDPC} code design, which raises the need for a design tool tailored to short length codes \cite{sEXIT}.
An optimal EXIT chart-based LDPC code design (e.g., matched EXIT curves) is inefficient for a short \ac{LDPC} code, as shown in Fig.~\ref{fig:shortlength}. One can see that a PEG-optimized short LDPC code realization following the asymptotically optimal degree profiles (see Fig.~\ref{fig:EXIT_matched}) has a worse error-rate performance, shown in Fig.~\ref{fig:BLER_36_GenAlg}, when compared to another short LDPC code designed by our proposed algorithm, and whose EXIT curves are not well-matched and even intersecting in the high mutual information region (see Fig.~\ref{fig:GenAlg_EXIT}). This reaffirms our aforementioned statements about the inefficiency of the classical code design tools in the short-length regime.

\section{Genetic Algorithm-based LDPC Code Design}
	\vspace{-0.05cm}
\begin{figure}[t]	
	\centering
	\resizebox{\columnwidth}{!}{
		\begin{tikzpicture}[rotate=-90, scale=.625]
\tikzset{
	edge/.style = {thick,black},
	mydiamond/.style={draw, diamond, aspect=1.5,text width=1.9cm, inner sep=0pt,  fill=white!90!red,text centered},
	BPrectangle/.style={rectangle, draw, minimum height=1cm, fill=white!80!gray, fill opacity=0.9,text centered},
	BPrectanglew/.style={rectangle, draw, minimum height=1cm, fill=white, fill opacity=0.9,text centered}
}

\draw [inner sep=0pt,ultra thick, draw=none, fill=white!65!blue, fill opacity=0.9]
(6.25,7.5) --(8.25,7.5) -- node[below, xshift=0cm,yshift=-0.04cm, rotate=0] {\color{blue!40!blue}{Decoder-in-the-Loop}} (8.25,2.5) -- (6.25,2.5) -- cycle;

\node[BPrectangle,  minimum height=0.5cm, text width=3cm] (pop) at (-1, 5) {Initialize Population};
\draw [edge,->](-3,6)--node[right, xshift=0cm] {$m$} (-2.5,6)--(-2.5,5.25) -- ([yshift=0.25cm]pop.north);
\draw [edge,->](-2.625,5) to node[above, yshift=0.3cm] {$n$} (pop);
\draw [edge,->](-3,4)--node[left, xshift=0cm,yshift=-0cm] {$k$}(-2.5,4)--(-2.5,4.75)-- ([yshift=-0.25cm]pop.north);
\draw [edge,-]([yshift=0cm]pop.south) to node[right, yshift=0.05cm] {$\mathbb{P}_{1}$} (0.75,5);
\draw[black,fill=black] (0.75,5) circle (.5ex); 
\draw [edge,->]([yshift=0cm]pop.south) -- (0.5,5);

\node[BPrectanglew,  minimum height=0.5cm,text width=2.33cm] (enc) at (2.5, 5) {LDPC Encoder};

\draw [edge,->](1.25, 4.75   ) -- ([yshift=-0.25cm]enc.north);

\node[BPrectanglew, minimum height=0.5cm, text width=1.3cm] (ch) at (4.5, 5) {Channel};
\draw [edge,->]([yshift=0cm]enc.south) to node[right, xshift=0cm,yshift=0cm] {$\mathbf{x}$}([yshift=0cm]ch.north);

\node[BPrectanglew, fill=white!65!blue, text width=2.5cm] (dec) at (7.25, 5) {LDPC Decoder (e.g., BP, OSD)};
\draw [edge,->]([yshift=0cm]ch.south) --(5.25,5) to node[right,xshift=0cm,yshift=0cm] {$\mathbf{y}$}([yshift=0cm]dec.north);

\draw[black,fill=black] (1.25, 4.75  ) circle (.5ex); 

\node[BPrectangle, text width=3.75cm] (upd) at (6, -2) {Update Population (Mutations \& Crossovers)};

\draw [edge,->]([yshift=-0cm]dec.west) to node[below,yshift=0cm] {$\mathbf{BLERs}$} (7.25,-2)--([yshift=0cm]upd.south);

\node[mydiamond] (cond) at (3, -2) {$i<N_{pop,max}?$};
\draw [edge,->](upd) -- node[right,yshift=-0.1cm] {$\mathbb{P}_{i+1}$}  (cond);
\draw [edge,-](cond) -- node[right,red] {yes} (1.5,-2)--(0.75,-2)-- node[above] {$\mathbb{P}_{i+1}$}  (0.75,4.5)--(1.25, 4.75   );

\node[] at(-0.8,-3.5) {Population $i$ $\left(\mathbb{P}_{i}\right)$: set of candidate};
\node[] at(-0.3,-1.7) {LDPC codes};

\draw[black,fill=black] (0.75,4.5) circle (.5ex); 
\draw [edge,->](0.75,1)-- (0.75,3);

\draw [edge,->](1.1,4.5) to  [bend left ](1.1,5.5);

\node[BPrectangle, minimum height=0.5cm, text width=2.5cm] (select) at (4.5, -8.5) { Select fittest $\mathbf{H}$};
\node[draw,ellipse] (ter) at (7.5, -8.5) {Terminate};

\draw [edge,->](cond) --node[above,red] {no}  (3,-8.5)--(select.north);
\node[] at (3.5,-7.375) {$\mathbb{P}_{N_{pop,max}}$};
\draw [edge,->](select) -- node[right,xshift=0cm] {$\mathbf{H}_{GenAlg}$} (ter);

\end{tikzpicture}
	} 
	\vspace{-0.5cm}
	\caption{\footnotesize Abstract view of genetic algorithm (GenAlg)-based LDPC code design.}
	\label{fig:block-diagram-LDPC}  
	\vspace{-0.65cm}	
\end{figure}
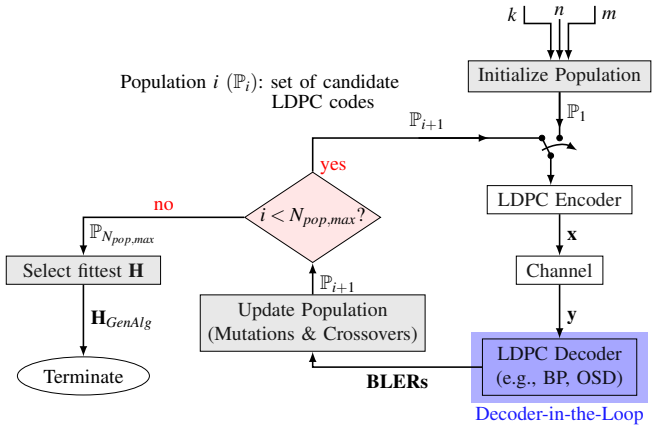

We consider the design of an LDPC code as an optimization problem, in which the target is to minimize the block error-rate (BLER) at a certain design SNR.  
This optimization problem has some constraints defined by the problem at hand. The code rate $R_c = 1 - \frac{\text{rank}\left( \mathbf{H} \right)}{n}$, the number of VNs and the number of CNs should remain constant. 
In other words, we keep the number of VNs fixed to $n$ (i.e., no puncturing involved) and the number of CNs fixed to $m$ (i.e., we assume no redundant checks\footnote{In case the resulting $\mathbf{H}$-matrix is not of full rank, it holds that $R_c > r_d$ which, if needed, could be (in a naive implementation) solved by freezing some VNs. Thus, we do not impose any further constraint other than $m$ CNs.}).
Furthermore, every variable (or check) node must be connected to at least one check (or variable) node, respectively.
To solve this problem, similar to \cite{GenAlg_Journal}, we apply the genetic algorithm (GenAlg) \cite{GeneticsFirstpaper}.

The design process starts with an initial population of some randomly constructed LDPC codes (i.e., \emph{population 1}).
An error-rate computation framework is used to assess the error-rate performance of the found LDPC codes at a certain design SNR and a fixed maximum number of BP iterations $N_{it,max}$.
The best LDPC codes from this population are picked and then undergo evolutionary transformations (mutations and crossovers).
This process is repeated until a certain target error-rate or a maximum number of populations (i.e., epochs) is reached, see Fig.~\ref{fig:block-diagram-LDPC}.

The ``mutations'' are done by adding (or removing) an edge to (or from) the parent $\mathbf{H}$-matrix at a random position, or a combination of both, Fig.~\ref{fig:mutation}.
The ``crossover'' is a symmetric 2D-crossover between two parent $\mathbf{H}$-matrices (i.e., $\mathbf{H}_1$ and $\mathbf{H}_2$). The left (or upper) half matrix of $\mathbf{H}_1$ is concatenated with the right (or lower) half matrix of $\mathbf{H}_2$) forming two offspring $\mathbf{H}$-matrices in the next population, Fig.~\ref{fig:crossover}.

Population $(i+1)$ contains the best (in terms of error-rate) $T$ $\mathbf{H}$-matrices from population $i$, together with mutated offsprings from those $T$ $\mathbf{H}$-matrices and offsprings due to crossover between each pair of $T$ parent $\mathbf{H}$-matrices. 
For all simulation results using the GenAlg as discussed next, we set $T = 20$.
We refer to \cite{GenAlg_Journal} for further details on the GenAlg for code design.
For the sake of reproducibility, the source code is available online.

\section{Insights From Optimized LDPC Codes Over AWGN Channel}

\vspace{-0.1cm}

To be consistent with the results in \cite{liva2016}, we design LDPC codes with code length $n=128$, code dimension $k=64$ and, thus, code rate $R_c=0.5$. All considered LDPC codes in this section are simulated over the binary-input AWGN (bi-AWGN) channel. All reference LDPC codes taken from \cite{liva2016} were designed through a girth optimization technique based on the \ac{PEG} algorithm and are considered state-of-the-art:
the standardized LDPC code by the Consultative Committee for Space Data Systems (CCSDS) for satellite telecommand links (\ref{lab:protonasa}),
an accumulate-repeat-3-accumulate (AR3A) LDPC code (\ref{lab:ara}),
an accumulate-repeat-jagged-accumulate (ARJA) LDPC code~(\ref{lab:arja}),
and the proposed protograph-based LDPC code for the upcoming 5G NR standard with a base graph (base graph 2 in \cite{polar5G2018}) optimized for small blocklengths (\ref{lab:5G}).
As a calibration step of our decoding framework, we were able to reproduce exactly the same BLER curves using our own simulation setup. Therefore, the presented gains are not an artefact of different decoder implementations.

To get started, our initial population $\mathbb{P}_{1}$ contains a set of randomly constructed regular $\left( 3,6 \right)$ LDPC codes (no \ac{PEG} used).

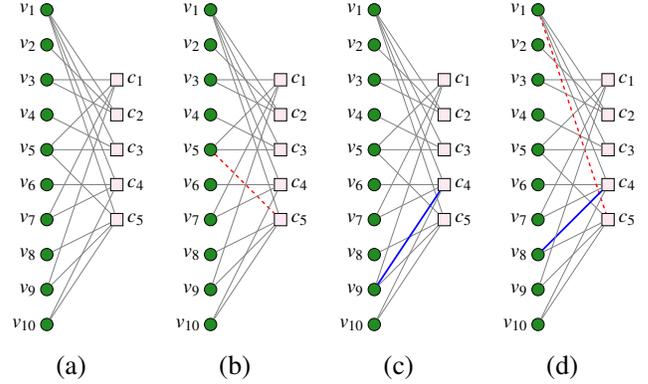
\begin{figure}[t]
	\resizebox{0.999\columnwidth}{!}{
		
		\begin{subfigure}{0.2\columnwidth}
			\resizebox{\columnwidth}{!}{\tikzset{text=black, font={\fontsize{16pt}{10}\selectfont}}
\let\pgfmathMod=\pgfmathmod\relax
\definecolor{lightgrey}{gray}{0.5}

\begin{tikzpicture}[rotate=\angleRot]
\tikzset{edge/.style = {-, thick}}

\tikzset{h1/.style={preaction={draw,yellow,-,double=yellow,double distance=4\pgflinewidth,}}}

\tikzset{h2/.style={preaction={draw,green,-,double=green,double distance=4\pgflinewidth,}}}

\node [cnds, label=\cndPos:{$c_{1}$}](c1) at(3,5) {};

\node [cnds, label=\cndPos:{$c_{2}$}](c2) at(3,4) {};

\node [cnds, label=\cndPos:{$c_{3}$}](c3) at(3,3) {};

\node [cnds, label=\cndPos:{$c_{4}$}](c4) at(3,2) {};

\node [cnds, label=\cndPos:{$c_{5}$}](c5) at(3,1) {};

\node [vndGs, label=\vndPos:{$v_{1}$}](v1) at(1,7) {};

\node [vndGs, label=\vndPos:{$v_{2}$}](v2) at(1,6) {};
\node [vndGs, label=\vndPos:{$v_{3}$}](v3) at(1,5) {};
\node [vndGs, label=\vndPos:{$v_{4}$}](v4) at(1,4) {};
\node [vndGs, label=\vndPos:{$v_{5}$}](v5) at(1,3) {};
\node [vndGs, label=\vndPos:{$v_{6}$}](v6) at(1,2) {};
\node [vndGs, label=\vndPos:{$v_{7}$}](v7) at(1,1) {};
\node [vndGs, label=\vndPos:{$v_{8}$}](v8) at(1,0) {};
\node [vndGs, label=\vndPos:{$v_{9}$}](v9) at(1,-1) {};
\node [vndGs, label=\vndPos:{$v_{10}$}](v10) at(1,-2) {};

\draw[lightgrey,line width=0.25mm] (c1)--(v3);
\draw[lightgrey,line width=0.25mm] (c1)--(v7);

\draw[lightgrey,line width=0.25mm] (c2)--(v2);

\draw[lightgrey,line width=0.25mm] (c3)--(v4);

\draw[lightgrey,line width=0.25mm] (c4)--(v6);

\draw[lightgrey,line width=0.25mm] (c5)--(v8);

\draw[lightgrey,line width=0.25mm] (v1)--(c3);
\draw[lightgrey,line width=0.25mm] (v1)--(c5);
\draw[lightgrey,line width=0.25mm] (v5)--(c5);

\draw[lightgrey,line width=0.25mm] (c2)--(v1);
\draw[lightgrey,line width=0.25mm] (c1)--(v5);
\draw[lightgrey,line width=0.25mm] (c4)--(v1);
\draw[lightgrey,line width=0.25mm] (c1)--(v9);
\draw[lightgrey,line width=0.25mm] (c2)--(v3);
\draw[lightgrey,line width=0.25mm] (c3)--(v5);
\draw[lightgrey,line width=0.25mm] (c4)--(v7);
\draw[lightgrey,line width=0.25mm] (c5)--(v9);

\draw[lightgrey,line width=0.25mm] (c4)--(v10);
\draw[lightgrey,line width=0.25mm] (c5)--(v10);

\end{tikzpicture}}
			\caption{\small}
			\label{fig:bipartite1}
		\end{subfigure}
		\begin{subfigure}{0.2\columnwidth}
			\resizebox{\columnwidth}{!}{\tikzset{text=black, font={\fontsize{16pt}{10}\selectfont}}
\let\pgfmathMod=\pgfmathmod\relax
\definecolor{lightgrey}{gray}{0.5}

\begin{tikzpicture}[rotate=\angleRot]
\tikzset{edge/.style = {-, thick}}

\tikzset{h1/.style={preaction={draw,yellow,-,double=yellow,double distance=4\pgflinewidth,}}}

\tikzset{h2/.style={preaction={draw,green,-,double=green,double distance=4\pgflinewidth,}}}

\node [cnds, label=\cndPos:{$c_{1}$}](c1) at(3,5) {};

\node [cnds, label=\cndPos:{$c_{2}$}](c2) at(3,4) {};

\node [cnds, label=\cndPos:{$c_{3}$}](c3) at(3,3) {};

\node [cnds, label=\cndPos:{$c_{4}$}](c4) at(3,2) {};

\node [cnds, label=\cndPos:{$c_{5}$}](c5) at(3,1) {};

\node [vndGs, label=\vndPos:{$v_{1}$}](v1) at(1,7) {};

\node [vndGs, label=\vndPos:{$v_{2}$}](v2) at(1,6) {};
\node [vndGs, label=\vndPos:{$v_{3}$}](v3) at(1,5) {};
\node [vndGs, label=\vndPos:{$v_{4}$}](v4) at(1,4) {};
\node [vndGs, label=\vndPos:{$v_{5}$}](v5) at(1,3) {};
\node [vndGs, label=\vndPos:{$v_{6}$}](v6) at(1,2) {};
\node [vndGs, label=\vndPos:{$v_{7}$}](v7) at(1,1) {};
\node [vndGs, label=\vndPos:{$v_{8}$}](v8) at(1,0) {};
\node [vndGs, label=\vndPos:{$v_{9}$}](v9) at(1,-1) {};
\node [vndGs, label=\vndPos:{$v_{10}$}](v10) at(1,-2) {};

\draw[lightgrey,line width=0.25mm] (c1)--(v3);
\draw[lightgrey,line width=0.25mm] (c1)--(v7);

\draw[lightgrey,line width=0.25mm] (c2)--(v2);

\draw[lightgrey,line width=0.25mm] (c3)--(v4);

\draw[lightgrey,line width=0.25mm] (c4)--(v6);

\draw[lightgrey,line width=0.25mm] (c5)--(v8);

\draw[lightgrey,line width=0.25mm] (v1)--(c3);
\draw[lightgrey,line width=0.25mm] (v1)--(c5);
\draw[dashed, line width=0.4mm, red] (v5)--(c5);

\draw[lightgrey,line width=0.25mm] (c2)--(v1);
\draw[lightgrey,line width=0.25mm] (c1)--(v5);
\draw[lightgrey,line width=0.25mm] (c4)--(v1);
\draw[lightgrey,line width=0.25mm] (c1)--(v9);
\draw[lightgrey,line width=0.25mm] (c2)--(v3);
\draw[lightgrey,line width=0.25mm] (c3)--(v5);
\draw[lightgrey,line width=0.25mm] (c4)--(v7);
\draw[lightgrey,line width=0.25mm] (c5)--(v9);

\draw[lightgrey,line width=0.25mm] (c4)--(v10);
\draw[lightgrey,line width=0.25mm] (c5)--(v10);
\end{tikzpicture}}
			\caption{\small}
			\label{fig:bipartite2}
		\end{subfigure}
		\begin{subfigure}{0.2\columnwidth}
			\resizebox{\columnwidth}{!}{\tikzset{text=black, font={\fontsize{16pt}{10}\selectfont}}
\let\pgfmathMod=\pgfmathmod\relax
\definecolor{lightgrey}{gray}{0.5}

\begin{tikzpicture}[rotate=\angleRot]
\tikzset{edge/.style = {-, thick}}

\tikzset{h1/.style={preaction={draw,yellow,-,double=yellow,double distance=4\pgflinewidth,}}}

\tikzset{h2/.style={preaction={draw,green,-,double=green,double distance=4\pgflinewidth,}}}

\node [cnds, label=\cndPos:{$c_{1}$}](c1) at(3,5) {};

\node [cnds, label=\cndPos:{$c_{2}$}](c2) at(3,4) {};

\node [cnds, label=\cndPos:{$c_{3}$}](c3) at(3,3) {};

\node [cnds, label=\cndPos:{$c_{4}$}](c4) at(3,2) {};

\node [cnds, label=\cndPos:{$c_{5}$}](c5) at(3,1) {};

\node [vndGs, label=\vndPos:{$v_{1}$}](v1) at(1,7) {};

\node [vndGs, label=\vndPos:{$v_{2}$}](v2) at(1,6) {};
\node [vndGs, label=\vndPos:{$v_{3}$}](v3) at(1,5) {};
\node [vndGs, label=\vndPos:{$v_{4}$}](v4) at(1,4) {};
\node [vndGs, label=\vndPos:{$v_{5}$}](v5) at(1,3) {};
\node [vndGs, label=\vndPos:{$v_{6}$}](v6) at(1,2) {};
\node [vndGs, label=\vndPos:{$v_{7}$}](v7) at(1,1) {};
\node [vndGs, label=\vndPos:{$v_{8}$}](v8) at(1,0) {};
\node [vndGs, label=\vndPos:{$v_{9}$}](v9) at(1,-1) {};
\node [vndGs, label=\vndPos:{$v_{10}$}](v10) at(1,-2) {};

\draw[lightgrey,line width=0.25mm] (c1)--(v3);
\draw[lightgrey,line width=0.25mm] (c1)--(v7);

\draw[lightgrey,line width=0.25mm] (c2)--(v2);

\draw[lightgrey,line width=0.25mm] (c3)--(v4);

\draw[lightgrey,line width=0.25mm] (c4)--(v6);

\draw[lightgrey,line width=0.25mm] (c5)--(v8);

\draw[lightgrey,line width=0.25mm] (v1)--(c3);
\draw[lightgrey,line width=0.25mm] (v1)--(c5);
\draw[lightgrey,line width=0.25mm] (v5)--(c5);

\draw[lightgrey,line width=0.25mm] (c2)--(v1);
\draw[lightgrey,line width=0.25mm] (c1)--(v5);
\draw[lightgrey,line width=0.25mm] (c4)--(v1);
\draw[lightgrey,line width=0.25mm] (c1)--(v9);
\draw[lightgrey,line width=0.25mm] (c2)--(v3);
\draw[lightgrey,line width=0.25mm] (c3)--(v5);
\draw[lightgrey,line width=0.25mm] (c4)--(v7);
\draw[lightgrey,line width=0.25mm] (c5)--(v9);
\draw[line width=0.5mm, blue] (c4)--(v9);

\draw[lightgrey,line width=0.25mm] (c4)--(v10);
\draw[lightgrey,line width=0.25mm] (c5)--(v10);

\end{tikzpicture}}
			\caption{\small}
			\label{fig:bipartite3}
		\end{subfigure}
		\begin{subfigure}{0.2\columnwidth}
			\resizebox{\columnwidth}{!}{\tikzset{text=black, font={\fontsize{16pt}{10}\selectfont}}
\let\pgfmathMod=\pgfmathmod\relax
\definecolor{lightgrey}{gray}{0.5}

\begin{tikzpicture}[rotate=\angleRot]
\tikzset{edge/.style = {-, thick}}

\tikzset{h1/.style={preaction={draw,yellow,-,double=yellow,double distance=4\pgflinewidth,}}}

\tikzset{h2/.style={preaction={draw,green,-,double=green,double distance=4\pgflinewidth,}}}

\node [cnds, label=\cndPos:{$c_{1}$}](c1) at(3,5) {};

\node [cnds, label=\cndPos:{$c_{2}$}](c2) at(3,4) {};

\node [cnds, label=\cndPos:{$c_{3}$}](c3) at(3,3) {};

\node [cnds, label=\cndPos:{$c_{4}$}](c4) at(3,2) {};

\node [cnds, label=\cndPos:{$c_{5}$}](c5) at(3,1) {};

\node [vndGs, label=\vndPos:{$v_{1}$}](v1) at(1,7) {};

\node [vndGs, label=\vndPos:{$v_{2}$}](v2) at(1,6) {};
\node [vndGs, label=\vndPos:{$v_{3}$}](v3) at(1,5) {};
\node [vndGs, label=\vndPos:{$v_{4}$}](v4) at(1,4) {};
\node [vndGs, label=\vndPos:{$v_{5}$}](v5) at(1,3) {};
\node [vndGs, label=\vndPos:{$v_{6}$}](v6) at(1,2) {};
\node [vndGs, label=\vndPos:{$v_{7}$}](v7) at(1,1) {};
\node [vndGs, label=\vndPos:{$v_{8}$}](v8) at(1,0) {};
\node [vndGs, label=\vndPos:{$v_{9}$}](v9) at(1,-1) {};
\node [vndGs, label=\vndPos:{$v_{10}$}](v10) at(1,-2) {};

\draw[lightgrey,line width=0.25mm] (c1)--(v3);
\draw[lightgrey,line width=0.25mm] (c1)--(v7);

\draw[lightgrey,line width=0.25mm] (c2)--(v2);

\draw[lightgrey,line width=0.25mm] (c3)--(v4);

\draw[lightgrey,line width=0.25mm] (c4)--(v6);

\draw[lightgrey,line width=0.25mm] (c5)--(v8);

\draw[lightgrey,line width=0.25mm] (v1)--(c3);
\draw[dashed, line width=0.4mm, red] (v1)--(c5);
\draw[lightgrey,line width=0.25mm] (v5)--(c5);

\draw[line width=0.5mm, blue] (c4)--(v8);

\draw[lightgrey,line width=0.25mm] (c2)--(v1);
\draw[lightgrey,line width=0.25mm] (c1)--(v5);
\draw[lightgrey,line width=0.25mm] (c4)--(v1);
\draw[lightgrey,line width=0.25mm] (c1)--(v9);
\draw[lightgrey,line width=0.25mm] (c2)--(v3);
\draw[lightgrey,line width=0.25mm] (c3)--(v5);
\draw[lightgrey,line width=0.25mm] (c4)--(v7);
\draw[lightgrey,line width=0.25mm] (c5)--(v9);

\draw[lightgrey,line width=0.25mm] (c4)--(v10);
\draw[lightgrey,line width=0.25mm] (c5)--(v10);

\end{tikzpicture}}
			\caption{\small}
			\label{fig:bipartite4}
		\end{subfigure}
	} \vspace{-0.25cm}\caption{\footnotesize Mutation examples; All derived from parent (a) by: removing an edge (b), adding an edge (c), or a combination of both (d).}
	\vspace{-0.4cm} \label{fig:mutation}
\end{figure}

\begin{figure}[t]
	\begin{center}
		\resizebox{\columnwidth}{!}{\begin{tikzpicture}

\node[] at (-2.3,1) {
	$
	\left[ \begin{array}{llllllll} 
	
	\color{red}1&	\color{red}1&	\color{red}0&	\color{red}1&	\color{darkgreen}1&	\color{darkgreen}1&	\color{darkgreen}0&	\color{darkgreen}1 \\
	\color{red}1&	\color{red}0&	\color{red}1&	\color{red}1&	\color{darkgreen}1&	\color{darkgreen}0&	\color{darkgreen}1&	\color{darkgreen}1 \\
	\color{blue}1&	\color{blue}1&	\color{blue}1&	\color{blue}0&	\color{cyan}1&	\color{cyan}1&	\color{cyan}1&	\color{cyan}0 \\
	\color{blue}0&	\color{blue}1&	\color{blue}1&	\color{blue}1&	\color{cyan}0&	\color{cyan}1&	\color{cyan}1&	\color{cyan}1 \\
	
	\end{array}\right] 
	$ 
	};

\node[] at (2.3,1) {
$
\left[ \begin{array}{llllllll} 

\color{black}1&	\color{black}1&	\color{black}1&	\color{black}0&	\color{magenta}1&	\color{magenta}0&	\color{magenta}1&	\color{magenta}1 \\
\color{black}0&	\color{black}1&	\color{black}1&	\color{black}1&	\color{magenta}1&	\color{magenta}1&	\color{magenta}0&	\color{magenta}1 \\
\color{brown}1&	\color{brown}1&	\color{brown}0&	\color{brown}1&	\color{gray}1&	\color{gray}1&	\color{gray}1&	\color{gray}0 \\
\color{brown}1&	\color{brown}0&	\color{brown}1&	\color{brown}1&	\color{gray}0&	\color{gray}1&	\color{gray}1&	\color{gray}1 \\

\end{array}\right] 
$ 
	};

\node[] at (-2.3,-1.5) {
$
\left[ \begin{array}{llllllll} 

\color{red}1&	\color{red}1&	\color{red}0&	\color{red}1&	\color{magenta}1&	\color{magenta}0&	\color{magenta}1&	\color{magenta}1 \\
\color{red}1&	\color{red}0&	\color{red}1&	\color{red}1&	\color{magenta}1&	\color{magenta}1&	\color{magenta}0&	\color{magenta}1 \\
\color{blue}1&	\color{blue}1&	\color{blue}1&	\color{blue}0&	\color{gray}1&	\color{gray}1&	\color{gray}1&	\color{gray}0 \\
\color{blue}0&	\color{blue}1&	\color{blue}1&	\color{blue}1&	\color{gray}0&	\color{gray}1&	\color{gray}1&	\color{gray}1 \\

\end{array}\right] 
$ 
	};
	
\node[] at (2.3,-1.5) {
$
\left[ \begin{array}{llllllll} 

\color{red}1&	\color{red}1&	\color{red}0&	\color{red}1&	\color{darkgreen}1&	\color{darkgreen}1&	\color{darkgreen}0&	\color{darkgreen}1 \\
\color{red}1&	\color{red}0&	\color{red}1&	\color{red}1&	\color{darkgreen}1&	\color{darkgreen}0&	\color{darkgreen}1&	\color{darkgreen}1 \\
\color{brown}1&	\color{brown}1&	\color{brown}0&	\color{brown}1&	\color{gray}1&	\color{gray}1&	\color{gray}1&	\color{gray}0 \\
\color{brown}1&	\color{brown}0&	\color{brown}1&	\color{brown}1&	\color{gray}0&	\color{gray}1&	\color{gray}1&	\color{gray}1 \\

\end{array}\right] 
$ 
	};
	
	\node[] at (-2.2,-0.1) {(a) Parent 1 $\mathbf{H}_1$};
	\node[] at (2.3,-0.1) {(b) Parent 2 $\mathbf{H}_2$};	
	
	\node[] at (-2.2,-2.6) {(c) Offspring 1};
	\node[] at (2.3,-2.6) {(d) Offspring 2};
	
 \draw[draw=red, fill=red, opacity=0.2] (-4.25,1.05) rectangle (-2.4,1.85);		
 \draw[draw=red, fill=red, opacity=0.2] (-4.25,-0.6) rectangle (-2.4,-1.45);		 
 \draw[draw=red, fill=red, opacity=0.2] (0.3,-0.6) rectangle (2.15,-1.45);		  
 \draw[draw=blue, fill=blue, opacity=0.2] (-4.25,0.2) rectangle (-2.4,1);		 
 \draw[draw=blue, fill=blue, opacity=0.2] (-4.25,-1.5) rectangle (-2.4,-2.3);		  
 \draw[draw=gray, fill=gray, opacity=0.2, pattern=north west lines, pattern color=gray] (-2.2,-1.5) rectangle (-0.25,-2.3);		   
 \draw[draw=magenta, fill=magenta, opacity=0.2, pattern=north west lines, pattern color=magenta] (-2.2,-0.6) rectangle (-0.25,-1.45);		    
 \draw[draw=brown, fill=brown, opacity=0.2, pattern=north west lines, pattern color=brown] (0.3,-1.5) rectangle (2.15,-2.3);		   
 \draw[draw=gray, fill=gray, opacity=0.2, pattern=north west lines, pattern color=gray] (2.45,-1.5) rectangle (4.3,-2.3);		    
 \draw[draw=gray, fill=gray, opacity=0.2, pattern=north west lines, pattern color=gray] (2.45,0.2) rectangle (4.3,1);		    
 \draw[draw=magenta, fill=magenta, opacity=0.2, pattern=north west lines, pattern color=magenta] (2.45,1.05) rectangle (4.3,1.85);		  
 \draw[draw=brown, fill=brown, opacity=0.2, pattern=north west lines, pattern color=brown] (0.3,0.2) rectangle (2.15,1);		    
 \draw[draw=green, fill=green, opacity=0.2] (-2.2,1.05) rectangle (-0.25,1.85);		 
 \draw[draw=green, fill=green, opacity=0.2] (2.45,-0.6) rectangle (4.3,-1.45);		   
\end{tikzpicture}}
	\end{center}
\vspace{-0.5cm}
	\caption{\footnotesize Crossover examples between the two parents $\mathbf{H}_1$ and $\mathbf{H}_2$.}\label{fig:crossover}
\vspace{-0.55cm}
\end{figure}
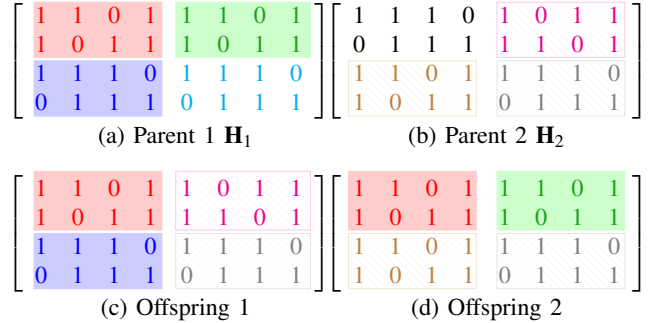

\vspace{-0.1cm}

\subsection{Error-rate performance}

Using \ac{GenAlg}, we inherently design the whole edge interleaver (i.e., $\mathbf{H}$-matrix) tailored to BP decoding with a maximum number of BP iterations $N_{it,max}=200$ at a design SNR of $5$ dB. The resulting LDPC code (\ref{plot:GenAlg_200_5dB}) performs equally good as the 5G LDPC code over the whole simulated SNR range, as shown in Fig. \ref{fig:64128LDPCBP200}.

\begin{figure}[H]
   	\vspace{-0.3cm}	
	\captionsetup[subfigure]{justification=centering}
	
	\begin{subfigure}{\columnwidth}
		\begin{tikzpicture}
\begin{semilogyaxis}[
     width=\linewidth,
     height=0.75\linewidth,
     grid style={dotted,anthrazit},
     xmajorgrids,
     yminorticks=false,
     ymajorgrids,
     legend cell align=left,
       legend style={font=\footnotesize},
       legend columns=1,
       legend style={at={(0.001,0.3)},anchor=west,draw=none,
       	/tikz/every even column/.append style={column sep=0.5cm}
       },
	xmin=2,
	xmax=5.63,
	ymin=1e-6,
	ytick={1e-6,1e-4,1e-2,1e0},
	ymax=0.335,
	xlabel={$E_b/N_0$ [dB]},
	ylabel={BLER},
	legend image post style={mark indices={}},
	mark options={solid}
	]

\addplot[color=black,line width = 0.5pt,solid,mark=x]
table[row sep=crcr]{
	1	                0.769650000000000\\
	1.50000000000000	0.561760000000000\\
	2	                0.334810000000000\\
	2.50000000000000	0.151430000000000\\
	3	                0.0496700000000000\\
	3.50000000000000	0.0115768000000000\\
	4	                0.00187240000000000\\
	4.50000000000000	0.000209666666666667\\
	5	                1.68400000000000e-05\\
	5.50000000000000	1.05000000000000e-06\\
};
\label{lab:protonasa}

\addplot[color=blue,line width = 0.5pt,solid,mark=triangle]
table[row sep=crcr]{
	1	 0.584660000000000\\
	1.5	 0.366480000000000\\
	2	 0.185100000000000\\
	2.5	 0.0712400000000000\\
	3	 0.0203700000000000\\
	3.5	 0.00449180000000000\\
	4	 0.000769400000000000\\
	4.5	 0.000113000000000000\\
	5	 1.46400000000000e-05\\
	5.5	 1.61000000000000e-06\\
	5.75 7.12000000000000e-07\\
};
\label{lab:36}

\addplot[color=ahmedpurple,line width = 0.5pt,solid,mark=diamond] table[x=snr,y=cer] {./ARA.txt}; \label{lab:ara}  

\addplot[color=gray,line width = 0.5pt,solid,mark=square] table[x=snr,y=cer] {./ARJA.txt}; \label{lab:arja} 

\addplot [color=red,line width = 0.3pt,solid,mark=*,mark indices={1,5,9},mark size=2.5pt]
table[row sep=crcr]{
	1	0.464370000000000\\
	1.5	0.262120000000000\\
	2	0.114480000000000\\
	2.5	0.0380000000000000\\
	3	0.00903000000000000\\
	3.5	0.00160840000000000\\
	4	0.000208000000000000\\
	4.5	2.47000000000000e-05\\
	5	2.49000000000000e-06\\
	5.5	2.50000000000000e-07\\
};
\label{plot:GenAlg_200_5dB}

\addplot [color=red,line width = 0.3pt,solid,mark=square*,mark indices={2,6,9},mark size=2.5pt]
table[row sep=crcr]{
	1	0.455990000000000\\
	1.5	0.253450000000000\\
	2	0.109880000000000\\
	2.5	0.0349200000000000\\
	3	0.00800000000000000\\
	3.5	0.00145520000000000\\
	4	0.000200800000000000\\
	4.5	2.19666666666667e-05\\
	5	2.84000000000000e-06\\
	5.5	3.80000000000000e-07\\
};
\label{plot:RA_200}

\addplot [color=red,line width = 0.3pt,solid,mark=triangle*,mark indices={3,7,9},mark size=3.5pt]
table[row sep=crcr]{
	1	0.459980000000000\\
	1.5	0.258950000000000\\
	2	0.111380000000000\\
	2.5	0.0367500000000000\\
	3	0.00789000000000000\\
	3.5	0.00149540000000000\\
	4	0.000208800000000000\\
	4.5	2.52333333333333e-05\\
	5	2.97000000000000e-06\\
	5.5	4.10000000000000e-07\\
};
\label{plot:IRA_deg2_200}

\addplot [color=red,line width = 0.3pt,solid,mark=diamond*,mark indices={4,8,9},mark size=3.5pt]
table[row sep=crcr]{
	1	0.464080000000000\\
	1.5	0.260670000000000\\
	2	0.114440000000000\\
	2.5	0.0366600000000000\\
	3	0.00861000000000000\\
	3.5	0.00149880000000000\\
	4	0.000203200000000000\\
	4.5	2.45333333333333e-05\\
	5	3.17000000000000e-06\\
	5.5	3.60000000000000e-07\\
};
\label{plot:IRA_deg3_200}

\addplot[color=darkgreen,line width = 0.5pt,solid,mark=otimes,mark size=3.5pt]
table[row sep=crcr]{
0.750000000000000	0.567090000000000\\
1.25000000000000	0.352520000000000\\
1.75000000000000	0.172680000000000\\
2.25000000000000	0.0640700000000000\\
2.75000000000000	0.0178000000000000\\
3.25000000000000	0.00368120000000000\\
3.75000000000000	0.000538000000000000\\
4.25000000000000	6.68666666666667e-05\\
4.75000000000000	6.57666666666667e-06\\
5.25000000000000	5.80000000000000e-07\\
5.50000000000000	1.63000000000000e-07\\
};
\label{lab:5G}

\draw [red,very thick] (4.65,1.2e-5) ellipse (0.16cm and 0.07cm);
\node[red] at (3.5,1.8e-6) {GenAlg $\left(N_{it,max}=200\right)$};
\draw[->,-triangle 45,red,thick] (4.565,1.2e-5) -- (3.5,2.7e-6);

\coordinate (legend3) at (axis description cs:1.06,0.97);

\end{semilogyaxis}

\matrix [
draw=none,
matrix of nodes,
anchor=south east,
font=\footnotesize,
mark options={solid}
] at (legend3) {
	& Construction @ design SNR & & Construction @ design SNR \\
	\ref{lab:protonasa} &  CCSDS Up-Link LDPC \cite{NorbertWehn} & \ref{lab:36} &  (3,6) Regular LDPC \\
	\ref{lab:ara} &  AR3A LDPC from \cite{liva2016} &	\ref{lab:arja} &  ARJA LDPC from \cite{liva2016} \\
	\ref{lab:5G} &  5G LDPC \cite{polar5G2018} &	& \\	
	
	\ref{plot:GenAlg_200_5dB} &  GenAlg LDPC @ 5 dB & \ref{plot:RA_200} &  GenAlg IRA @ 5 dB \\	
	\ref{plot:IRA_deg2_200} &  GenAlg TB-IRA @ 5 dB & \ref{plot:IRA_deg3_200} &  GenAlg PTB-IRA @ 5 dB \\		
};
\end{tikzpicture}
		\vspace{-0.7cm}
		\caption{BLER comparison}
		\label{fig:64128LDPCBP200}
	\end{subfigure}
	
	\vspace{0.09cm}	
	
	\begin{subfigure}{\columnwidth}
		\begin{tikzpicture}
\begin{axis}[
     width=9cm,
     height=6.5cm,
     grid=both,
     grid style={dotted,anthrazit},
       legend cell align=left,
       legend style={font=\footnotesize},
       legend columns=1,
       legend style={at={(0.001,0.3)},anchor=west,draw=none,
       	/tikz/every even column/.append style={column sep=0.5cm}
       },
	xmin=2,
	xmax=4.5,
	xtick={1,1.5,2,2.5,3,3.5,4,4.5},
	ymin=0,
	ymax=72.5,
	ytick={0,20,40,60,80},	
	major tick length=0pt,
	xlabel={$E_b/N_0$ [dB]},
	ylabel={Average Iterations $N_{it,avg}$},
	legend image post style={mark indices={}},
	mark options={solid}
	]

\addplot[color=black,line width = 0.5pt,solid,mark=x]
table[row sep=crcr]{
	1	156.431030000000\\
	1.5	116.398930000000\\
	2	72.2425800000000\\
	2.5	35.9674600000000\\
	3	14.8927900000000\\
	3.5	5.98380960000000\\
	4	2.99233900000000\\
	4.5	2.02144406666667\\
	5	1.59729589000000\\
	5.5	1.34221494000000\\
	};
	\label{lab:protonasa} 

\addplot[color=blue,line width = 0.5pt,solid,mark=triangle]
table[row sep=crcr]{
1	121.002260000000\\
1.5	78.6135700000000\\
2	42.8941800000000\\
2.5	19.6942700000000\\
3	8.34283000000000\\
3.5	4.03507300000000\\
4	2.50530800000000\\
4.5	1.87608763333333\\
5	1.52692892000000\\
5.5	1.30041944000000\\
5.75	1.21908328800000\\
};
\label{lab:36}

\addplot [color=red,line width = 0.3pt,solid,mark=*,mark indices={1,5},mark size=2.5pt]
table[row sep=crcr]{
	1	96.4027200000000\\
	1.5	57.2253200000000\\
	2	28.4058300000000\\
	2.5	12.5701900000000\\
	3	5.83552000000000\\
	3.5	3.45902460000000\\
	4	2.57119260000000\\
	4.5	2.11007773333333\\
	5	1.78231662000000\\
	5.5	1.52766223000000\\
};
\label{plot:GenAlg_200_5dB1}

\addplot [color=red,line width = 0.3pt,solid,mark=square*,mark indices={2,6},mark size=2.5pt]
table[row sep=crcr]{
1	94.8105200000000\\
1.5	55.8268000000000\\
2	27.6939200000000\\
2.5	12.1139900000000\\
3	5.70782000000000\\
3.5	3.44059820000000\\
4	2.57889200000000\\
4.5	2.11499946666667\\
5	1.78362226000000\\
5.5	1.52392973000000\\
};
\label{plot:RA_2001}

\addplot [color=red,line width = 0.3pt,solid,mark=triangle*,mark indices={3},mark size=3.5pt]
table[row sep=crcr]{
1	95.7743000000000\\
1.5	56.8989000000000\\
2	28.0069400000000\\
2.5	12.4494000000000\\
3	5.69264000000000\\
3.5	3.44825520000000\\
4	2.57359200000000\\
4.5	2.10827493333333\\
5	1.77738261000000\\
5.5	1.51890045000000\\
};
\label{plot:IRA_deg2_2001}

\addplot [color=red,line width = 0.3pt,solid,mark=diamond*,mark indices={4},mark size=3.5pt]
table[row sep=crcr]{
1	96.4866400000000\\
1.5	57.2597900000000\\
2	28.5421800000000\\
2.5	12.4559700000000\\
3	5.83469000000000\\
3.5	3.45078660000000\\
4	2.57198480000000\\
4.5	2.10748360000000\\
5	1.77666653000000\\
5.5	1.51807067000000\\
};
\label{plot:IRA_deg3_2001}

\addplot[color=darkgreen,line width = 0.5pt,solid,mark=otimes,mark size=3.5pt]
table[row sep=crcr]{
0.750000000000000	118.390840000000\\
1.25000000000000	77.7213000000000\\
1.75000000000000	43.1078300000000\\
2.25000000000000	20.9579200000000\\
2.75000000000000	10.1956600000000\\
3.25000000000000	5.93427420000000\\
3.75000000000000	4.35945140000000\\
4.25000000000000	3.66831206666667\\
4.75000000000000	3.24822504333333\\
5.25000000000000	2.94198719000000\\
5.50000000000000	2.81417049850000\\
};
\label{lab:5G}

\coordinate (legend) at (axis description cs:1,0.155);
\end{axis}

\matrix [
draw,
fill=white,
matrix of nodes,
anchor=south east,
font=\footnotesize,
mark options={solid}
] at (legend) {
	& Construction @ design SNR \\
	\ref{lab:protonasa} &  CCSDS Up-Link LDPC \cite{NorbertWehn} \\
	\ref{lab:36} &  (3,6) Regular LDPC \\
	\ref{lab:5G} &  5G LDPC \cite{polar5G2018} \\	
	\ref{plot:GenAlg_200_5dB} &  GenAlg LDPC @ 5 dB \\
	\ref{plot:RA_200} &  GenAlg IRA @ 5 dB \\
	\ref{plot:IRA_deg2_200} & GenAlg TB-IRA @ 5 dB \\	
	\ref{plot:IRA_deg3_200} &  GenAlg PTB-IRA @ 5 dB \\		
};

\end{tikzpicture}
		\vspace{-0.7cm}
		\caption{Average number of required iterations}
		\label{fig:64128LDPC_avgIter200}
	\end{subfigure}
	
	\vspace{0.18cm}	
	
	\begin{subfigure}{\columnwidth}
		\begin{tikzpicture}
\begin{axis}[
     width=9cm,
     height=7cm,
     grid=both,
     grid style={dotted,anthrazit},
       legend cell align=left,
       legend style={font=\footnotesize},
       legend columns=1,
       legend style={at={(0.001,0.3)},anchor=west,draw=none,
       	/tikz/every even column/.append style={column sep=0.5cm}
       },
	xmin=2,
	xmax=4.5,
	xtick={1,1.5,2,2.5,3,3.5,4,4.5},
	ymin=0,
	ymax=72.5*512/64,
	ytick={0,150,300,450,600},	
	y tick label style={/pgf/number format/.cd,
		set thousands separator={}},
	major tick length=0pt,
	xlabel={$E_b/N_0$ [dB]},
	ylabel={Average decoding complexity $\eta$},			
	legend image post style={mark indices={}},
	mark options={solid}
	]

\addplot[color=black,line width = 0.5pt,solid,mark=x]
table[y expr=\thisrowno{1}*512/64, x expr=\thisrowno{0}, row sep=crcr]{
	1	156.431030000000\\
	1.5	116.398930000000\\
	2	72.2425800000000\\
	2.5	35.9674600000000\\
	3	14.8927900000000\\
	3.5	5.98380960000000\\
	4	2.99233900000000\\
	4.5	2.02144406666667\\
	5	1.59729589000000\\
	5.5	1.34221494000000\\
	};
	\label{lab:protonasa} 

\addplot[color=blue,line width = 0.5pt,solid,mark=triangle]
table[y expr=\thisrowno{1}*384/64, x expr=\thisrowno{0}, row sep=crcr]{%
1	121.002260000000\\
1.5	78.6135700000000\\
2	42.8941800000000\\
2.5	19.6942700000000\\
3	8.34283000000000\\
3.5	4.03507300000000\\
4	2.50530800000000\\
4.5	1.87608763333333\\
5	1.52692892000000\\
5.5	1.30041944000000\\
5.75	1.21908328800000\\
};
\label{lab:36}

\addplot [color=red,line width = 0.3pt,solid,mark=*,mark indices={1,5},mark size=2.5pt]
table[y expr=\thisrowno{1}*419/64, x expr=\thisrowno{0}, row sep=crcr]{%
	1	96.4027200000000\\
	1.5	57.2253200000000\\
	2	28.4058300000000\\
	2.5	12.5701900000000\\
	3	5.83552000000000\\
	3.5	3.45902460000000\\
	4	2.57119260000000\\
	4.5	2.11007773333333\\
	5	1.78231662000000\\
	5.5	1.52766223000000\\
};
\label{plot:GenAlg_200_5dB1}

\addplot [color=red,line width = 0.3pt,solid,mark=square*,mark indices={2,6},mark size=2.5pt]
table[y expr=\thisrowno{1}*437/64, x expr=\thisrowno{0}, row sep=crcr]{%
1	94.8105200000000\\
1.5	55.8268000000000\\
2	27.6939200000000\\
2.5	12.1139900000000\\
3	5.70782000000000\\
3.5	3.44059820000000\\
4	2.57889200000000\\
4.5	2.11499946666667\\
5	1.78362226000000\\
5.5	1.52392973000000\\
};
\label{plot:RA_2001}

\addplot [color=red,line width = 0.3pt,solid,mark=triangle*,mark indices={3},mark size=3.5pt]
table[y expr=\thisrowno{1}*438/64, x expr=\thisrowno{0}, row sep=crcr]{%
1	95.7743000000000\\
1.5	56.8989000000000\\
2	28.0069400000000\\
2.5	12.4494000000000\\
3	5.69264000000000\\
3.5	3.44825520000000\\
4	2.57359200000000\\
4.5	2.10827493333333\\
5	1.77738261000000\\
5.5	1.51890045000000\\
};
\label{plot:IRA_deg2_2001}

\addplot [color=red,line width = 0.3pt,solid,mark=diamond*,mark indices={4},mark size=3.5pt]
table[y expr=\thisrowno{1}*439/64, x expr=\thisrowno{0}, row sep=crcr]{%
1	96.4866400000000\\
1.5	57.2597900000000\\
2	28.5421800000000\\
2.5	12.4559700000000\\
3	5.83469000000000\\
3.5	3.45078660000000\\
4	2.57198480000000\\
4.5	2.10748360000000\\
5	1.77666653000000\\
5.5	1.51807067000000\\
};
\label{plot:IRA_deg3_2001}

\addplot[color=darkgreen,line width = 0.5pt,solid,mark=otimes,mark size=3.5pt]
table[y expr=\thisrowno{1}*473/64, x expr=\thisrowno{0}, row sep=crcr]{%
0.750000000000000	118.390840000000\\
1.25000000000000	77.7213000000000\\
1.75000000000000	43.1078300000000\\
2.25000000000000	20.9579200000000\\
2.75000000000000	10.1956600000000\\
3.25000000000000	5.93427420000000\\
3.75000000000000	4.35945140000000\\
4.25000000000000	3.66831206666667\\
4.75000000000000	3.24822504333333\\
5.25000000000000	2.94198719000000\\
5.50000000000000	2.81417049850000\\
};
\label{lab:5G}

\coordinate (legend) at (axis description cs:1,0.233);

\node[black,text opacity=1, draw, fill=white, fill opacity=1] at (2.52,530) {\small $\eta = \frac{N_{it,avg} \cdot E}{k}$};

\end{axis}

\matrix [
draw,
fill=white,
matrix of nodes,
anchor=south east,
font=\footnotesize,
mark options={solid}
] at (legend) {
	& Construction @ design SNR \\
	\ref{lab:protonasa} &  CCSDS Up-Link LDPC \cite{NorbertWehn} \\
	\ref{lab:36} &  (3,6) Regular LDPC \\
	\ref{lab:5G} &  5G LDPC \cite{polar5G2018} \\	
	\ref{plot:GenAlg_200_5dB} &  GenAlg LDPC @ 5 dB \\
	\ref{plot:RA_200} &  GenAlg IRA @ 5 dB \\
	\ref{plot:IRA_deg2_200} & GenAlg TB-IRA @ 5 dB \\	
	\ref{plot:IRA_deg3_200} &  GenAlg PTB-IRA @ 5 dB \\		
};

\end{tikzpicture}
		\vspace{-0.7cm}
		\caption{Average decoding complexity; $E$ is the number of edges in the graph of the code}
		\label{fig:AWGN_64128LDPC_avgComp200}
	\end{subfigure}
	\vspace{-0.1cm}	
	\caption{\footnotesize Several $\left(n=128,k=64\right)$ LDPC codes decoded with BP decoding using \underline{$N_{it,max}=200$ iterations} over the \underline{bi-AWGN channel}.}
	\label{fig:AWGN_200_ALL}
\end{figure}

To facilitate the encoding of the \ac{GenAlg}-based LDPC codes, we design accumulator-based codes (i.e., a structured interleaver). We refer the interested reader to \cite{LivaStructuredLDPC,shuLin} for more details about different structured types and design methods of LDPC codes.  
In this work, we design \ac{IRA} codes\footnote{Similar to the LDPC codes of the DVB-S.2 standard.} such that the $\mathbf{H}$-matrix has the form

\vspace{-0.7cm}

\begin{align*} 
	\mathbf{H} = \left[ \mathbf{H}_L \ \ \mathbf{H}_R \right] 
\end{align*}
\vspace{-0.1cm}
where $\mathbf{H}_L$ is the sub-matrix to be optimized and $\mathbf{H}_R$ is a fixed ``dual-diagonal'' square sub-matrix with dimensions $m \times m$ 

\vspace{-0.2cm}

\begin{align*} 
\mathbf{H}_R^\textrm{{IRA}} = \left[ \begin{array}{lllllll} 
1 &  &  &  & & & 
\\1 & 1 &  & 
\\ & 1 & \ddots & & 
\\ &  & \ddots & \ddots & 	
\\ & &  & \ddots & 1
\\ &  & &  & 1 & 1
\\ &  & &  &  & 1 & 1
\end{array}\right]. 
\end{align*}
The resulting \ac{IRA} code without tailbiting (\ref{plot:RA_200}) designed at $\unit[5]{dB}$ under 200 BP iterations approaches, again, the performance of the 5G LDPC code, as shown in Fig.~\ref{fig:64128LDPCBP200}.

To avoid degree-1 VNs, an upper-right $1$ is included in the $\mathbf{H}$-matrix for \ac{TB-IRA} codes, such that $\mathbf{H}_R$ has the form 
\begin{align*} 
\mathbf{H}_R^{\textrm{TB-IRA}} = \left[ \begin{array}{lllllll} 
1 &  &  &  & & & \textcolor{rot}{1}
\\1 & 1 &  & 
\\ & 1 & \ddots & & 
\\ &  & \ddots & \ddots & 	
\\ & &  & \ddots & 1
\\ &  & &  & 1 & 1
\\ &  & &  &  & 1 & 1
\end{array}\right]. 
\end{align*}
The \ac{TB-IRA} code has only degree-$2$ VNs in the $\mathbf{H}_R$-matrix. The GenAlg-based \ac{TB-IRA} code (\ref{plot:IRA_deg2_200}) designed at $5$ dB again approaches the performance of the 5G LDPC code, as shown in Fig.~\ref{fig:64128LDPCBP200}.

To facilitate the encoding step, while still avoiding degree-1 VNs, a weight-three column replaces the weight-one column in the \ac{IRA} code and is then moved to the first column in $\mathbf{H}_R$, and thus the name \ac{PTB-IRA} codes, as 
\begin{align*} 
	\mathbf{H}_R^{\textrm{PTB-IRA}} = \left[ \begin{array}{lllllll} 
		\textcolor{rot}{1} & 1 &  &  &  &  
		\\ & 1 & 1 &  & 
		\\  & & 1 & \ddots & & 
		\\ \textcolor{rot}{1} & &  & \ddots & \ddots &  	
		\\  & & &  & \ddots & 1
		\\ & &  & &  & 1 & 1
		\\ \textcolor{rot}{1} & &  & &  &  & 1  
	\end{array}\right]. 
\end{align*}
This code is similar to the WiMAX LDPC codes (IEEE 802.16e) and the WiFi LDPC codes (IEEE 802.11n). The resulting GenAlg-based \ac{PTB-IRA} code (\ref{plot:IRA_deg3_200}) designed at $\unit[5]{dB}$ again approaches the performance of the 5G LDPC code, as shown in Fig.~\ref{fig:64128LDPCBP200}.

In some applications, it is crucial to ensure that the worst case decoding latency is relatively low while having an acceptable error-rate performance. 
In BP decoding, the worst case latency is proportional to $N_{it,max}$. 
So we design LDPC codes tailored to iterative \ac{BP} decoding with reduced maximum number of \ac{BP} iterations $N_{it,max}$. 
Using \ac{GenAlg}, an LDPC code tailored to $N_{it,max}=20$ \ac{BP} iterations (\ref{plot:GenAlg_20}) designed at $\unit[5]{dB}$ outperforms the 5G LDPC code, leading to an $E_b/N_0$ gain of $\unit[0.325]{dB}$ at BLER of $10^{-5}$, as shown in Fig. \ref{fig:64128LDPCBP20}. 
The resulting code under 20 BP iterations (\ref{plot:GenAlg_20}) approaches the error-rate performance of the LDPC codes from \cite{liva2016}, with lower $N_{it,max}$. For example, when compared to an AR3A LDPC code decoded with $N_{it,max}=200$ iterations (\ref{plot:ARA_outlier}). Thus, the proposed code can be decoded with reduced worst case decoding latency with competitive error-rate performance.

It is fair to mention that the 5G LDPC code was designed to support a wide range of blocklengths and code rates. Thus, the error-rate performance was not the only design target.
The 5G LDPC codes enables high degree of implementation parallelism and an organized message passing process, besides being described in a compact manner \cite{Richardson5G}.
Fortunately, other such structural constraints can be imposed in our genetic optimization problem to further simplify encoding and decoding implementations of the resulting GenAlg-based LDPC codes.

	\vspace{-0.1cm}
	
\subsection{Decoding latency and complexity}

We show that for a fixed $N_{it,max}$, significant decoding latency and decoding complexity reduction can be achieved only by optimizing the LDPC code edge interleaver.
In BP decoding of LDPC codes, the average decoding latency can be measured using the average number of BP iterations needed by the decoder $N_{it,avg}$. Due to the early stopping condition used, the average number of performed iterations $N_{it,avg}$ is much lower than $N_{it,max}$, especially in the high SNR region. Fig.~\ref{fig:64128LDPC_avgIter200} and Fig.~\ref{fig:64128LDPC_avgIter20} show that our proposed GenAlg-based LDPC codes required on average a lower number of BP iterations when compared to conventionally designed (reference) LDPC codes for the same $N_{it,max}$. This potentially leads to a reduction in the decoding latency. Thus, higher throughput decoder implementations are possible.

In a parallel iterative (message passing) decoder, the decoding complexity depends on the number of performed iterations and the number of arithmetic operations per iteration.
The number of arithmetic operations per iteration is proportional to the number of edges in the Tanner graph of the code $E$.
In other words, the decoding complexity heavily depends on the total number of messages passed between the VNs and the CNs. 
This means that an average decoding complexity measure can be calculated as the product of the average number of iterations $N_{it,avg}$ (due to the early stopping condition) and the number of edges in the Tanner graph $E$.

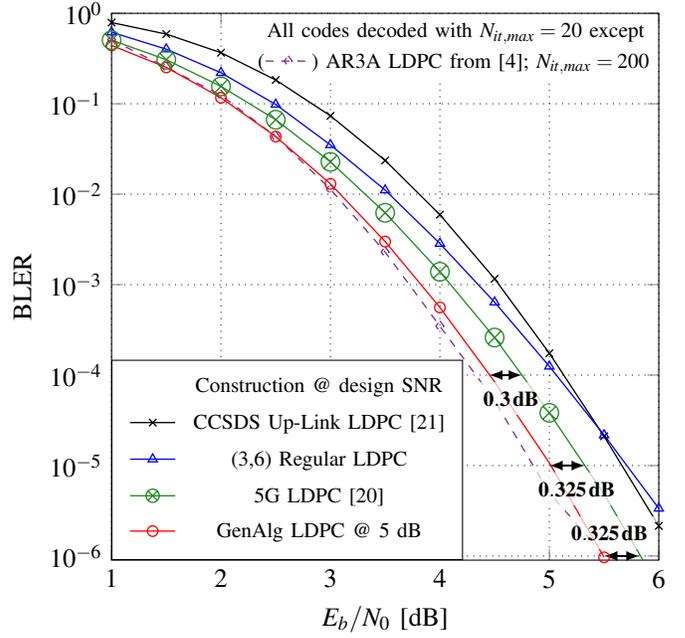
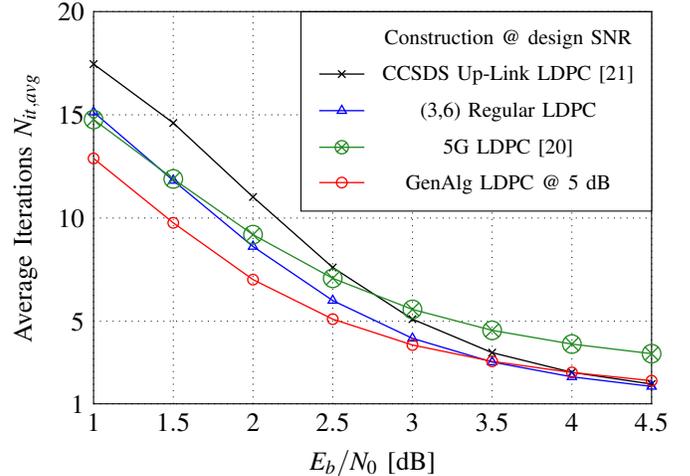
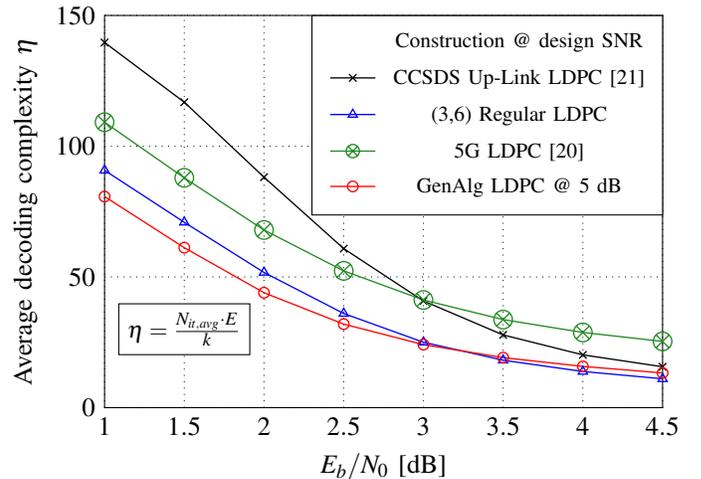
\begin{figure}[H]
	\vspace{-0.2cm}
	\captionsetup[subfigure]{justification=centering}
	
	\begin{subfigure}{\columnwidth}
		\begin{tikzpicture}
\begin{semilogyaxis}[
     width=\linewidth,
     height=\linewidth,
     grid style={dotted,anthrazit},
     xmajorgrids,
     yminorticks=false,
     ymajorgrids,
       legend cell align=left,
       legend style={font=\footnotesize},
       legend columns=1,
       legend style={at={(0.001,0.3)},anchor=west,draw=none,
       	/tikz/every even column/.append style={column sep=0.5cm}
       },
	xmin=1,
	xmax=6,
	ymin=9e-7,
	ymax=1,
	xlabel={$E_b/N_0$ [dB]},
	ylabel={BLER},
	legend image post style={mark indices={}},
	]

\addplot[color=ahmedpurple,line width = 0.5pt,dashed,mark=diamond,mark indices={1,2,3,4,5,6,7}] table[x=snr,y=cer] {./ARA.txt}; \label{plot:ARA_outlier}

\node[black,text opacity=1,fill=white, fill opacity=1] at (4.145,6e-1) {\footnotesize All codes decoded with $N_{it,max}=20$ except};
\node[black,text opacity=1,fill=white, fill opacity=1] at (4.125,2.8e-1) {\footnotesize (\ref{plot:ARA_outlier}) AR3A LDPC from \cite{liva2016}; $N_{it,max}=200$};

\addplot [color=black,line width = 0.5pt,solid,mark=x]
table[row sep=crcr]{
	1	0.790250000000000\\
	1.5	0.589220000000000\\
	2	0.366900000000000\\
	2.5	0.183320000000000\\
	3	0.0735300000000000\\
	3.5	0.0236162000000000\\
	4	0.00593640000000000\\
	4.5	0.00116230000000000\\
	5	0.000174490000000000\\
	5.5	2.10500000000000e-05\\
	6	2.15000000000000e-06\\
};
\label{plot:CCSDS_20}

\addplot[color=blue,line width = 0.5pt,solid,mark=triangle]
table[row sep=crcr]{%
	1	0.618180000000000\\
	1.5	0.401630000000000\\
	2	0.220160000000000\\
	2.5	0.0982500000000000\\
	3	0.0351000000000000\\
	3.5	0.0110928000000000\\
	4	0.00284500000000000\\
	4.5	0.000639200000000000\\
	5	0.000125040000000000\\
	5.5	2.17300000000000e-05\\
	6	3.36000000000000e-06\\
};
\label{lab:36_20}

\addplot [color=red,line width = 0.5pt,solid,mark=o,mark indices={1,2,3,4,5,6,7,10}]
table[row sep=crcr]{%
	1	0.441860000000000\\
	1.5	0.252420000000000\\
	2	0.116520000000000\\
	2.5	0.0434000000000000\\
	3	0.0130200000000000\\
	3.5	0.00299340000000000\\
	4	0.000560200000000000\\
	4.5	8.43666666666667e-05\\
	5	1.07000000000000e-05\\
	5.5	9.66666666666667e-07\\
};
\label{plot:GenAlg_20}

\addplot[color=darkgreen,line width = 0.5pt,solid,mark=otimes,mark size=3.5pt,mark indices={1,2,3,4,5,6,7,8,9,11}]
table[row sep=crcr]{
	1	                0.505020000000000\\
	1.50000000000000	0.307190000000000\\
	2	                0.155810000000000\\
	2.50000000000000	0.0666100000000000\\
	3	                0.0227800000000000\\
	3.50000000000000	0.00622220000000000\\
	4	                0.00138840000000000\\
	4.50000000000000	0.000260166666666667\\
	5	                3.82800000000000e-05\\
	5.50000000000000	4.82000000000000e-06\\
	6	                4.55833333333333e-07\\
};
\label{lab:5G_20}

\coordinate (legend3) at (axis description cs:0.6423,0);

\draw[<->,black,line width=0.75pt] (4.45,1e-4) -- (4.75,1e-4);
\node[black,text opacity=1,fill=white, fill opacity=0.7] at (4.65,5.5e-5) {\footnotesize $\mathbf{\unit[0.3]{\mathbf{dB}}}$};

\draw[<->,black,line width=0.75pt] (5,1e-5) -- (5.325,1e-5);
\node[black,text opacity=1,fill=white, fill opacity=0.7] at (5.25,5.5e-6) {\footnotesize $\mathbf{\unit[0.325]{\mathbf{dB}}}$};

\draw[<->,black,line width=0.75pt] (5.5,1e-6) -- (5.825,1e-6);
\node[black,text opacity=1,fill=white, fill opacity=0.7] at (5.55,1.8e-6) {\footnotesize $\mathbf{\unit[0.325]{\mathbf{dB}}}$};

\end{semilogyaxis}

\matrix [
draw,
fill=white,
matrix of nodes,
anchor=south east,
font=\footnotesize,
mark options={solid}
] at (legend3) {
	& Construction @ design SNR \\
	\ref{plot:CCSDS_20} &  CCSDS Up-Link LDPC \cite{NorbertWehn} \\	
	\ref{lab:36_20} &  (3,6) Regular LDPC \\	
	\ref{lab:5G_20} &  5G LDPC \cite{polar5G2018} \\	
	\ref{plot:GenAlg_20} &  GenAlg LDPC @ 5 dB \\
};

\end{tikzpicture}
		\vspace{-0.7cm}
		\caption{BLER comparison}
		\label{fig:64128LDPCBP20}
	\end{subfigure}
	
	\vspace{0.07cm}
	
	\begin{subfigure}{\columnwidth}
		\begin{tikzpicture}
\begin{axis}[
     width=9cm,
     height=6.8cm,
     grid=both,
     grid style={dotted,anthrazit},
       legend cell align=left,
       legend style={font=\footnotesize},
       legend columns=1,
       legend style={at={(0.001,0.3)},anchor=west,draw=none,
       	/tikz/every even column/.append style={column sep=0.5cm}
       },
	xmin=1,
	xmax=4.5,
	xtick={1,1.5,2,2.5,3,3.5,4,4.5},
	ymin=1,
	ymax=20,
	ytick={1,5,10,15,20},	
	major tick length=0pt,
	xlabel={$E_b/N_0$ [dB]},
	ylabel={Average Iterations $N_{it,avg}$},
	legend image post style={mark indices={}},
	mark options={solid}
	]

\addplot [color=black,line width = 0.5pt,solid,mark=x]
table[row sep=crcr]{%
1	17.4581000000000\\
1.5	14.6029500000000\\
2	11.0189100000000\\
2.5	7.60404000000000\\
3	5.10127000000000\\
3.5	3.47803760000000\\
4	2.52487440000000\\
4.5	1.95600290000000\\
5	1.59044113000000\\
5.5	1.34165472000000\\
6	1.17759988000000\\
};
\label{plot:CCSDS_20}

\addplot[color=blue,line width = 0.5pt,solid,mark=triangle]
table[row sep=crcr]{%
1	15.1279400000000\\
1.5	11.8168600000000\\
2	8.61951000000000\\
2.5	6.00299000000000\\
3	4.17074000000000\\
3.5	3.02409840000000\\
4	2.30650860000000\\
4.5	1.84119153333333\\
5	1.52148941000000\\
5.5	1.29979448000000\\
6	1.15487161000000\\
};
\label{lab:36_20}

\addplot [color=red,line width = 0.5pt,solid,mark=o]
table[row sep=crcr]{%
1	12.8858100000000\\
1.5	9.76613000000000\\
2	7.01571000000000\\
2.5	5.10111000000000\\
3	3.84923000000000\\
3.5	3.05569940000000\\
4	2.51819440000000\\
4.5	2.11858963333333\\
5	1.80206018333333\\
5.5	1.54881398333333\\
};
\label{plot:GenAlg_20}

\addplot[color=darkgreen,line width = 0.5pt,solid,mark=otimes,mark size=3.5pt]
table[row sep=crcr]{
	1	                14.7712500000000\\
	1.50000000000000	11.8932500000000\\
	2	                9.19466000000000\\
	2.50000000000000	7.08000000000000\\
	3	                5.57351000000000\\
	3.50000000000000	4.55834240000000\\
	4  	                3.89331900000000\\
	4.50000000000000	3.42905953333333\\
	5	                3.08421510000000\\
	5.50000000000000	2.81407900000000\\
	6	                2.59587139750000\\
};
\label{lab:5G_20}

\coordinate (legend) at (axis description cs:1.001,0.49);
\end{axis}

\matrix [
draw,
fill=white,
matrix of nodes,
anchor=south east,
font=\footnotesize,
mark options={solid}
] at (legend) {
	& Construction @ design SNR \\
	\ref{plot:CCSDS_20} &  CCSDS Up-Link LDPC \cite{NorbertWehn} \\	
	\ref{lab:36_20} &  (3,6) Regular LDPC \\	
	\ref{lab:5G_20} &  5G LDPC \cite{polar5G2018} \\	
	\ref{plot:GenAlg_20} &  GenAlg LDPC @ 5 dB \\
};

\end{tikzpicture}
		\vspace{-0.7cm}
		\caption{Average number of required iterations}
		\label{fig:64128LDPC_avgIter20}
	\end{subfigure}
	
	\vspace{0.07cm}
	
	\begin{subfigure}{\columnwidth}
		\begin{tikzpicture}
\begin{axis}[
     width=9cm,
     height=6.8cm,
     grid=both,
     grid style={dotted,anthrazit},
       legend cell align=left,
       legend style={font=\footnotesize},
       legend columns=1,
       legend style={at={(0.001,0.3)},anchor=west,draw=none,
       	/tikz/every even column/.append style={column sep=0.5cm}
       },
	xmin=1,
	xmax=4.5,
	xtick={1,1.5,2,2.5,3,3.5,4,4.5},
	ymin=0,
	ymax=150,
	major tick length=0pt,
	xlabel={$E_b/N_0$ [dB]},
	ylabel={Average decoding complexity $\eta$},		
	legend image post style={mark indices={}},
	mark options={solid}
	]

\addplot [color=black,line width = 0.5pt,solid,mark=x]
table[y expr=\thisrowno{1}*512/64, x expr=\thisrowno{0}, row sep=crcr]{
1	17.4581000000000\\
1.5	14.6029500000000\\
2	11.0189100000000\\
2.5	7.60404000000000\\
3	5.10127000000000\\
3.5	3.47803760000000\\
4	2.52487440000000\\
4.5	1.95600290000000\\
5	1.59044113000000\\
5.5	1.34165472000000\\
6	1.17759988000000\\
};
\label{plot:CCSDS_20}

\addplot[color=blue,line width = 0.5pt,solid,mark=triangle]
table[y expr=\thisrowno{1}*384/64, x expr=\thisrowno{0}, row sep=crcr]{
1	15.1279400000000\\
1.5	11.8168600000000\\
2	8.61951000000000\\
2.5	6.00299000000000\\
3	4.17074000000000\\
3.5	3.02409840000000\\
4	2.30650860000000\\
4.5	1.84119153333333\\
5	1.52148941000000\\
5.5	1.29979448000000\\
6	1.15487161000000\\
};
\label{lab:36_20}

\addplot [color=red,line width = 0.5pt,solid,mark=o]
table[y expr=\thisrowno{1}*401/64, x expr=\thisrowno{0}, row sep=crcr]{%
1	12.8858100000000\\
1.5	9.76613000000000\\
2	7.01571000000000\\
2.5	5.10111000000000\\
3	3.84923000000000\\
3.5	3.05569940000000\\
4	2.51819440000000\\
4.5	2.11858963333333\\
5	1.80206018333333\\
5.5	1.54881398333333\\
};
\label{plot:GenAlg_20}

\addplot[color=darkgreen,line width = 0.5pt,solid,mark=otimes,mark size=3.5pt]
table[y expr=\thisrowno{1}*473/64, x expr=\thisrowno{0}, row sep=crcr]{
	1	                14.7712500000000\\
	1.50000000000000	11.8932500000000\\
	2	                9.19466000000000\\
	2.50000000000000	7.08000000000000\\
	3	                5.57351000000000\\
	3.50000000000000	4.55834240000000\\
	4  	                3.89331900000000\\
	4.50000000000000	3.42905953333333\\
	5	                3.08421510000000\\
	5.50000000000000	2.81407900000000\\
	6	                2.59587139750000\\
};
\label{lab:5G_20}

\coordinate (legend) at (axis description cs:1.001,0.49);

\node[black,text opacity=1, draw, fill=white, fill opacity=1] at (1.5,30) {\small $\eta = \frac{N_{it,avg} \cdot E}{k}$};

\end{axis}

\matrix [
draw,
fill=white,
matrix of nodes,
anchor=south east,
font=\footnotesize,
mark options={solid}
] at (legend) {
	& Construction @ design SNR \\
	\ref{plot:CCSDS_20} &  CCSDS Up-Link LDPC \cite{NorbertWehn} \\	
	\ref{lab:36_20} &  (3,6) Regular LDPC \\	
	\ref{lab:5G_20} &  5G LDPC \cite{polar5G2018} \\	
	\ref{plot:GenAlg_20} &  GenAlg LDPC @ 5 dB \\
};

\end{tikzpicture}
		\vspace{-0.7cm}
		\caption{Average decoding complexity; $E$ is the number of edges in the graph of the code}
		\label{fig:AWGN_64128LDPC_avgComp20}
	\end{subfigure}
	\vspace{-0.09cm}	
	\caption{\footnotesize Several $\left(n=128,k=64\right)$ LDPC codes decoded with BP decoding using \underline{$N_{it,max}=20$ iterations} over the \underline{bi-AWGN channel}.}
	\label{fig:AWGN_20_ALL}
\end{figure}

Similar to \cite{finiteIter}, we use the average decoding complexity per information bit $\eta$ as the decoding complexity measure throughout this work
$$\eta = \frac{N_{it,avg} \cdot E}{k}.$$

Fig.~\ref{fig:AWGN_64128LDPC_avgComp200} and Fig.~\ref{fig:AWGN_64128LDPC_avgComp20} show a decoding complexity comparison between our proposed LDPC codes and the 5G LDPC codes under BP decoding with $N_{it,max}=200$ and $N_{it,max}=20$, respectively. This means that decoding complexity reduction was possible by designing the LDPC code using GenAlg.

Although our design algorithm, which depends on error-rate simulations (Fig.~\ref{fig:epochs}), is more complex than most of the conventional design tools, our proposed codes can be decoded with (much) lower complexity (Fig.~\ref{fig:AWGN_64128LDPC_avgComp200} and Fig.~\ref{fig:AWGN_64128LDPC_avgComp20}). Thus, there is a trade-off between \emph{offline} (i.e., design) complexity and \emph{online} (i.e., decoding) complexity. However, one should keep in mind that the design is only done once while the decoding complexity applies to every later usage of the designed code.
Furthermore, our proposed framework can be potentially used to design LDPC codes with the aim of reducing the decoding latency and/or complexity with a slightly relaxed error-rate performance constraint.
	
\subsection{Minimum distance $d_{min}$}
	
Computing the minimum distance $d_{min}$ of LDPC codes can be formulated as an integer program (see equation (\ref{minDistance})), which can be solved by numerical optimization methods \cite{dminLDPC}:

\begin{equation}
\begin{aligned}
& {\text{      min}}
& & \sum_{i=1}^{n}x_i\\
& \text{subject to}
&& \mathbf{H} \mathbf{x} - 2 \mathbf{z} = \mathbf{0}, \\
&&& \sum_{i=1}^{n}x_i \ge 1 \\
\end{aligned}
\label{minDistance}
\end{equation}
where $\mathbf{x} \in \left \{ 0,1 \right \}^n$, $\mathbf{z} \in \mathbb{Z}^m$ and all operations are performed over integer numbers. $d_{min}$ is the value of the objective function $\sum_{i=1}^{n}x_i$ at the minimum.

\begin{figure}[t]
	\begin{center}
		\begin{tikzpicture}
\begin{axis}[
width=\linewidth,
height=5cm,
xmin=1,
xmax=307,
xlabel={Population index $i$},
xtick={1,50,100,150,200,250,300},  
ytick={1e-5,4e-5,7e-5,1e-4},  
extra y ticks={1e-5},
improved scaled ticks={y}{base 10:5},
ylabel={BLER @ $\unit[5]{dB}$},
legend style={at={(axis cs:290,11e-5)},anchor=north east},
]
\addplot[color=red,solid,line width=1.5pt] table[x=epoch,y=cer] {./epochs2.txt};
\addlegendentry{GenAlg epochs};

\addplot[color=darkgreen,dashed,line width=1.75pt]
table[row sep=crcr]{%
	1     3.82800000000000e-05\\
	400   3.82800000000000e-05\\
};
\addlegendentry{5G LDPC \cite{polar5G2018}};

\end{axis}
\end{tikzpicture}
	\end{center}
	\vspace{-0.4cm}
	\caption{\footnotesize Evolution of the BLER at design SNR ${E_b}/{N_0} = \unit[5]{dB}$; $\left(n=128,k=64\right)$ LDPC codes; BP decoding with $N_{it,max}=20$; bi-AWGN channel.}
	\label{fig:epochs}
	\vspace{-0.6cm}
\end{figure}
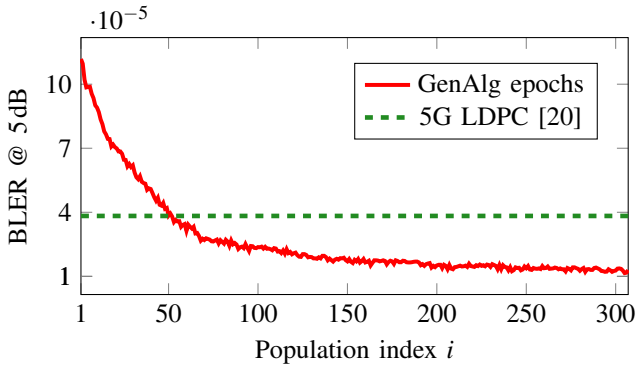

Table~\ref{tab:dmin} shows the $d_{min}$ of different LDPC codes. As a reference, we included the $d_{min}$ of the 5G polar code specified by the 3GPP group \cite{polar5G2018} and the \ac{RM} code with the same code length $n$ and code dimension $k$. The results in Table~\ref{tab:dmin}, Fig.~\ref{fig:64128LDPCBP200} and Fig.~\ref{fig:64128LDPCBP20} reassure that the minimum distance $d_{min}$ is not the only parameter to consider in order to enhance the performance of short linear codes under BP decoding. This can be attributed to the fact that the performance of a linear code under iterative decoding is dominated by the Tanner graph structure of the code and not its $d_{min}$ \cite{vardyStoppingDistance}. However, it is worth mentioning that maximizing $d_{min}$ is important to enhance the error-floor behavior of a code \cite{LivaStructuredLDPC}. The CCSDS LDPC code has the largest $d_{min}$ when compared to other LDPC codes considered in this work, because it was designed to operate in the very low error-rate region.

	\vspace{-0.25cm}
	
\begin{table}[h]
	\begin{center}
		\caption{\footnotesize $d_{min}$ of different $\left(n=128,k=64\right)$-codes}
		\label{tab:dmin}
			\vspace{-0.15cm}
		\begin{tabular}{ccc}
			\hline 
			Code & in Figures & $d_{min}$ \tabularnewline
			\hline 
			CCSDS Up-Link LDPC & \ref{lab:protonasa} & 14 \tabularnewline
			$(3,6)$ Regular LDPC & \ref{lab:36} & 8 \tabularnewline
			GenAlg LDPC @ 5 dB, 200 iter & \ref{plot:GenAlg_200_5dB} & 9 \tabularnewline
			GenAlg LDPC @ 5 dB, 20 iter & \ref{plot:GenAlg_20} & 8 \tabularnewline
			GenAlg IRA @ 5 dB& \ref{plot:RA_200} & 9 \tabularnewline
			GenAlg TB-IRA @ 5 dB& \ref{plot:IRA_deg2_200} & 9 \tabularnewline
			GenAlg PTB-IRA @ 5 dB& \ref{plot:IRA_deg3_200} & 9 \tabularnewline
			\hline 
			5G Polar (no CRC)&   & 8 \tabularnewline
			RM &   & 16 \tabularnewline
			\hline 
		\end{tabular}
	\end{center}
\vspace{-0.6cm}
\end{table}

\section{Results for the Rayleigh fading channel}

To demonstrate the flexibility of our proposed design algorithm, we also design LDPC codes for the Rayleigh fading channel.
To be more precise, we assume an ergodic Rayleigh fading model with full \ac{CSI} which can be motivated as the result of an \ac{OFDM}-based transmission in a multi-path propagation environment and, thus, is omnipresent in today’s wireless communication systems.
Our proposed design algorithm benefits from having the decoder-in-the-loop and the channel-in-the-loop. 
Thus, designing LDPC codes tailored to both the decoder and the channel is possible.

We design LDPC codes tailored to BP decoding with $N_{it,max}=200$ iterations at a design SNR $E_b/N_0 = \unit[8]{dB}$. As shown in Fig.~\ref{fig:Ray_BP200}, the resulting LDPC code (\ref{plot:Ray_GenAlg_200_8dB}) outperforms the 5G LDPC code optimized for short block lengths over the whole simulated SNR range (i.e., an $E_b/N_0$ gain of $\unit[0.2]{dB}$ at a BLER of $10^{-4}$). 
Moreover, our proposed LDPC code uses a lower number of iterations on average when compared to the 5G LDPC code as depicted in Fig.~\ref{fig:Ray_64128LDPC_avgIter200}. Also, the average decoding complexity needed to decode our proposed LDPC code is lower than that required for decoding the 5G LDPC code, see Fig.~\ref{fig:Ray_64128LDPC_avgComp200}. Thus, our proposed LDPC code has a better error-rate performance and can be decoded with reduced decoding latency and decoding complexity when compared to the 5G LDPC code.

Similarly, we use the same procedure to design LDPC codes tailored to a lower number of BP iterations $N_{it,max}=20$. In Fig.~\ref{fig:Ray_BP20}, we show that our GenAlg-designed LDPC code (\ref{plot:Ray_GenAlg_8dB_20}) outperforms the 5G LDPC code over the whole simulated SNR range and an $E_b/N_0$ gain of $\unit[0.8]{dB}$ is achieved at a BLER of $10^{-4}$. Again, a significant reduction in the average number of required iterations (i.e., decoding latency measure) and the average decoding complexity is achieved when compared to the reference (standardized) LDPC codes, see Fig.~\ref{fig:Ray_64128LDPC_avgIter20} and Fig.~\ref{fig:Ray_64128LDPC_avgComp20}. 

\begin{figure}[H]
	\vspace{-0.2cm}	
	\captionsetup[subfigure]{justification=centering}
	
	\begin{subfigure}{\columnwidth}
		\begin{tikzpicture}
\begin{axis}[
     width=\linewidth,
     height=\linewidth,
     grid style={dotted,anthrazit},
     xmajorgrids,
     yminorticks=false,
     ymajorgrids,
legend columns=1,
legend pos=south west,   
xlabel={$E_b/N_0$ [dB]},
ylabel={BLER},
ymode=log,
mark size=1.5pt,
xmin=1,
xmax=10,
ymin=2e-07,
ymax=1,
]

\addplot[color=black,line width = 0.5pt,solid,mark=x]
table[row sep=crcr]{
1	                0.996900000000000\\
1.90000000000000	0.968600000000000\\
2.90000000000000	0.813800000000000\\
3.90000000000000	0.507300000000000\\
4.90000000000000	0.186800000000000\\
5.90000000000000	0.0422000000000000\\
6.90000000000000	0.00518888888888889\\
7.90000000000000	0.000444500000000000\\
8.90000000000000	2.65500000000000e-05\\
9.90000000000000	1.05000000000000e-06\\
};
\label{lab:protonasa}

\addplot[color=blue,line width = 0.5pt,solid,mark=triangle]
table[row sep=crcr]{
1	                0.978100000000000\\
1.60000000000000	0.925100000000000\\
2.60000000000000	0.701600000000000\\
3.60000000000000	0.359000000000000\\
4.60000000000000	0.117700000000000\\
5.60000000000000	0.0246000000000000\\
6.60000000000000	0.00250000000000000\\
7.60000000000000	0.000237500000000000\\
8.60000000000000	1.91000000000000e-05\\
9.60000000000000	2.84000000000000e-06\\
10.6000000000000	5.68000000000000e-07\\
};
\label{lab:36}

\addplot[color=darkgreen,line width = 0.5pt,solid,mark=otimes,mark size=2.5pt]
table[row sep=crcr]{
1	                0.947900000000000\\
1.30000000000000	0.909000000000000\\
2.30000000000000	0.681800000000000\\
3.30000000000000	0.348200000000000\\
4.30000000000000	0.108200000000000\\
5.30000000000000	0.0205000000000000\\
6.30000000000000	0.00252000000000000\\
7.30000000000000	0.000235000000000000\\
8.30000000000000	1.66000000000000e-05\\
9.30000000000000	1.60000000000000e-06\\
10.3000000000000	1.20000000000000e-07\\
};
\label{lab:5G}

\addplot [color=red,line width = 0.5pt,solid,mark=*]
table[row sep=crcr]{
	1	0.945100000000000\\
	2	0.761700000000000\\
	3	0.435900000000000\\
	4	0.159700000000000\\
	5	0.0314000000000000\\
	6	0.00405555555555556\\
	7	0.000299500000000000\\
	8	2.10500000000000e-05\\
	9	1.81000000000000e-06\\
	10	1.70000000000000e-07\\
};
\label{plot:Ray_GenAlg_200_8dB}

\coordinate (legend) at (axis description cs:0.67,0.03);

\draw [black,line width = 1pt] (7.5,1e-4) ellipse (0.2cm and 0.08cm);
\node[black,text opacity=1,fill=white, fill opacity=0.7] at (7.6,5e-5) {\footnotesize $\mathbf{\unit[0.2]{\mathbf{dB}}}$};

\draw [black,line width = 1pt] (8.4,1e-5) ellipse (0.2cm and 0.08cm);
\node[black,text opacity=1,fill=white, fill opacity=0.7] at (8.4,5e-6) {\footnotesize $\mathbf{\unit[0.2]{\mathbf{dB}}}$};

\draw [black,line width = 1pt] (9.4,1e-6) ellipse (0.2cm and 0.08cm);
\node[black,text opacity=1,fill=white, fill opacity=0.7] at (9.4,5e-7) {\footnotesize $\mathbf{\unit[0.25]{\mathbf{dB}}}$};

\end{axis}

\matrix [
draw,
fill=white,
matrix of nodes,
anchor=south east,
font=\footnotesize,
mark options={solid}
] at (legend) {
	& Construction @ design SNR \\
	\ref{lab:protonasa} &  CCSDS Up-Link LDPC \cite{NorbertWehn} \\
	\ref{lab:36} &  (3,6) Regular LDPC \\
	\ref{lab:5G} &  5G LDPC \cite{polar5G2018} \\	
	\ref{plot:Ray_GenAlg_200_8dB} &  GenAlg LDPC @ 8 dB \\		
};

\end{tikzpicture}
		\vspace{-0.7cm}
		\caption{BLER comparison}
		\label{fig:Ray_BP200}
	\end{subfigure}
	
	\vspace{0.1cm}
	
	\begin{subfigure}{\columnwidth}
		\begin{tikzpicture}
\begin{axis}[
     width=9cm,
     height=6.7cm,
     grid=both,
     grid style={dotted,anthrazit},
       legend cell align=left,
       legend style={font=\footnotesize},
       legend columns=1,
       legend style={at={(0.001,0.3)},anchor=west,draw=none,
       	/tikz/every even column/.append style={column sep=0.5cm}
       },
	xmin=1,
	xmax=10,
	ymin=0,
	ymax=200,
	ytick={0,50,100,150,200},	
	major tick length=0pt,
	xlabel={$E_b/N_0$ [dB]},
	ylabel={Average Iterations $N_{it,avg}$},
	legend image post style={mark indices={}},
	mark options={solid}
	]

\addplot[color=black,line width = 0.5pt,solid,mark=x]
table[row sep=crcr]{
1	                199.441000000000\\
1.90000000000000	194.190600000000\\
2.90000000000000	165.025300000000\\
3.90000000000000	106.441800000000\\
4.90000000000000	43.9188000000000\\
5.90000000000000	14.1100000000000\\
6.90000000000000	4.70168888888889\\
7.90000000000000	2.55509000000000\\
8.90000000000000	1.90518020000000\\
9.90000000000000	1.57050354000000\\
	};
	\label{lab:Ray_protonasa_200} 

\addplot[color=blue,line width = 0.5pt,solid,mark=triangle]
table[row sep=crcr]{
1	                195.985100000000\\
1.60000000000000	186.135200000000\\
2.60000000000000	143.731300000000\\
3.60000000000000	77.7552000000000\\
4.60000000000000	30.0406000000000\\
5.60000000000000	9.81800000000000\\
6.60000000000000	3.81182222222222\\
7.60000000000000	2.42490300000000\\
8.60000000000000	1.88468910000000\\
9.60000000000000	1.56290461000000\\
10.6000000000000	1.34272489000000\\
};
\label{lab:Ray_36_200}

\addplot[color=darkgreen,line width = 0.5pt,solid,mark=otimes,mark size=2.5pt]
table[row sep=crcr]{
1	                190.512100000000\\
1.30000000000000	183.279500000000\\
2.30000000000000	140.756200000000\\
3.30000000000000	77.9933000000000\\
4.30000000000000	31.1461000000000\\
5.30000000000000	11.5706000000000\\
6.30000000000000	5.86194000000000\\
7.30000000000000	4.29269833333333\\
8.30000000000000	3.63358970000000\\
9.30000000000000	3.24532356500000\\
10.3000000000000	2.97145607900000\\
};
\label{lab:Ray_5G_200}

\addplot [color=red,line width = 0.5pt,solid,mark=*]
table[row sep=crcr]{
	1	189.388800000000\\
	3	91.3595000000000\\
	4	38.3281000000000\\
	5	11.6715000000000\\
	6	4.72470000000000\\
	7	3.00862450000000\\
	8	2.38713430000000\\
	9	2.00503005000000\\
	10	1.72100476200000\\
};
\label{plot:Ray_GenAlg_200_8dB}

\coordinate (legend) at (axis description cs:1,0.48);
\end{axis}

\matrix [
draw,
fill=white,
matrix of nodes,
anchor=south east,
font=\footnotesize,
mark options={solid}
] at (legend) {
	& Construction @ design SNR \\
	\ref{lab:Ray_protonasa_200} &  CCSDS Up-Link LDPC \cite{NorbertWehn} \\
	\ref{lab:Ray_36_200} &  (3,6) Regular LDPC \\
	\ref{lab:Ray_5G_200} &  5G LDPC \cite{polar5G2018} \\	
	\ref{plot:Ray_GenAlg_200_8dB} &  GenAlg LDPC @ 8 dB \\		
};

\end{tikzpicture}
		\vspace{-0.7cm}
		\caption{Average number of required iterations}
		\label{fig:Ray_64128LDPC_avgIter200}
	\end{subfigure}
	
	\vspace{0.3cm}
	
	\begin{subfigure}{\columnwidth}
		\begin{tikzpicture}
\begin{axis}[
     width=9cm,
     height=6.7cm,
     grid=both,
     grid style={dotted,anthrazit},
       legend cell align=left,
       legend style={font=\footnotesize},
       legend columns=1,
       legend style={at={(0.001,0.3)},anchor=west,draw=none,
       	/tikz/every even column/.append style={column sep=0.5cm}
       },
	xmin=1,
	xmax=10,
	ymin=0,
	ymax=1600,
	y tick label style={/pgf/number format/.cd,
    set thousands separator={}},
	major tick length=0pt,
	xlabel={$E_b/N_0$ [dB]},
	ylabel={Average decoding complexity $\eta$},	
	legend image post style={mark indices={}},
	mark options={solid}
	]

\addplot[color=black,line width = 0.5pt,solid,mark=x]
table[y expr=\thisrowno{1}*512/64, x expr=\thisrowno{0}, row sep=crcr]{
1	                199.441000000000\\
1.90000000000000	194.190600000000\\
2.90000000000000	165.025300000000\\
3.90000000000000	106.441800000000\\
4.90000000000000	43.9188000000000\\
5.90000000000000	14.1100000000000\\
6.90000000000000	4.70168888888889\\
7.90000000000000	2.55509000000000\\
8.90000000000000	1.90518020000000\\
9.90000000000000	1.57050354000000\\
	};
	\label{lab:Ray_protonasa_200} 

\addplot[color=blue,line width = 0.5pt,solid,mark=triangle]
table[y expr=\thisrowno{1}*384/64, x expr=\thisrowno{0}, row sep=crcr]{
1	                195.985100000000\\
1.60000000000000	186.135200000000\\
2.60000000000000	143.731300000000\\
3.60000000000000	77.7552000000000\\
4.60000000000000	30.0406000000000\\
5.60000000000000	9.81800000000000\\
6.60000000000000	3.81182222222222\\
7.60000000000000	2.42490300000000\\
8.60000000000000	1.88468910000000\\
9.60000000000000	1.56290461000000\\
10.6000000000000	1.34272489000000\\
};
\label{lab:Ray_36_200}

\addplot[color=darkgreen,line width = 0.5pt,solid,mark=otimes,mark size=2.5pt]
table[y expr=\thisrowno{1}*473/64, x expr=\thisrowno{0}, row sep=crcr]{
1	                190.512100000000\\
1.30000000000000	183.279500000000\\
2.30000000000000	140.756200000000\\
3.30000000000000	77.9933000000000\\
4.30000000000000	31.1461000000000\\
5.30000000000000	11.5706000000000\\
6.30000000000000	5.86194000000000\\
7.30000000000000	4.29269833333333\\
8.30000000000000	3.63358970000000\\
9.30000000000000	3.24532356500000\\
10.3000000000000	2.97145607900000\\
};
\label{lab:Ray_5G_200}

\addplot [color=red,line width = 0.5pt,solid,mark=*]
table[y expr=\thisrowno{1}*413/64, x expr=\thisrowno{0}, row sep=crcr]{
	1	189.388800000000\\
	2	153.910400000000\\
	3	91.3595000000000\\
	4	38.3281000000000\\
	5	11.6715000000000\\
	6	4.72470000000000\\
	7	3.00862450000000\\
	8	2.38713430000000\\
	9	2.00503005000000\\
	10	1.72100476200000\\
};
\label{plot:Ray_GenAlg_200_8dB}

\coordinate (legend) at (axis description cs:1,0.48);

\node[black,text opacity=1, draw, fill=white, fill opacity=1] at (2.25,300) {\small $\eta = \frac{N_{it,avg} \cdot E}{k}$};

\end{axis}

\matrix [
draw,
fill=white,
matrix of nodes,
anchor=south east,
font=\footnotesize,
mark options={solid}
] at (legend) {
	& Construction @ design SNR \\
	\ref{lab:Ray_protonasa_200} &  CCSDS Up-Link LDPC \cite{NorbertWehn} \\
	\ref{lab:Ray_36_200} &  (3,6) Regular LDPC \\
	\ref{lab:Ray_5G_200} &  5G LDPC \cite{polar5G2018} \\	
	\ref{plot:Ray_GenAlg_200_8dB} &  GenAlg LDPC @ 8 dB \\		
};

\end{tikzpicture}
		\vspace{-0.7cm}
		\caption{Average decoding complexity; $E$ is the number of edges in the graph of the code}
		\label{fig:Ray_64128LDPC_avgComp200}
	\end{subfigure}
	\vspace{-0.1cm}	
	\caption{\footnotesize Several $\left(n=128,k=64\right)$ LDPC codes decoded with BP decoding using \underline{$N_{it,max}=200$ iterations} over the \underline{Rayleigh fading channel}.}
	\label{fig:Ray_200_ALL}
\end{figure}

\begin{figure}[H]
	\vspace{-0.2cm}		
	\captionsetup[subfigure]{justification=centering}
	
	\begin{subfigure}{\columnwidth}
		\begin{tikzpicture}
\begin{axis}[
     width=\linewidth,
     height=\linewidth,
     grid style={dotted,anthrazit},
     xmajorgrids,
     yminorticks=false,
     ymajorgrids,
legend columns=1,
legend pos=south west,   
xlabel={$E_b/N_0$ [dB]},
ylabel={BLER},
ymode=log,
mark size=1.5pt,
xmin=1,
xmax=10,
ymin=4e-06,
ymax=1,
legend image post style={mark indices={}},
mark options={solid}
]

\addplot [color=black,line width = 0.5pt,solid,mark=x]
table[row sep=crcr]{
	1	                0.998200000000000\\
	1.90000000000000	0.973100000000000\\
	2.90000000000000	0.847100000000000\\
	3.90000000000000	0.531000000000000\\
	4.90000000000000	0.235800000000000\\
	5.90000000000000	0.0710000000000000\\
	6.90000000000000	0.0160555555555556\\
	7.90000000000000	0.00272750000000000\\
	8.90000000000000	0.000332550000000000\\
	9.90000000000000	2.98400000000000e-05\\
	10.9000000000000	2.01400000000000e-06\\
};
\label{plot:CCSDS_20}

\addplot[color=blue,line width = 0.5pt,solid,mark=triangle]
table[row sep=crcr]{
	1	                0.984200000000000\\
	1.60000000000000	0.939400000000000\\
	2.60000000000000	0.734700000000000\\
	3.60000000000000	0.412700000000000\\
	4.60000000000000	0.163300000000000\\
	5.60000000000000	0.0448000000000000\\
	6.60000000000000	0.00873333333333333\\
	7.60000000000000	0.00146500000000000\\
	8.60000000000000	0.000200000000000000\\
	9.60000000000000	2.28000000000000e-05\\
	10.6000000000000	2.10000000000000e-06\\
};
\label{lab:36_20}

\addplot[color=darkgreen,line width = 0.5pt,solid,mark=otimes,mark size=2.5pt]
table[row sep=crcr]{
	1	                0.963900000000000\\
	1.30000000000000	0.934600000000000\\
	2.30000000000000	0.728400000000000\\
	3.30000000000000	0.411400000000000\\
	4.30000000000000	0.171300000000000\\
	5.30000000000000	0.0484000000000000\\
	6.30000000000000	0.0107000000000000\\
	7.30000000000000	0.00191000000000000\\
	8.30000000000000	0.000269400000000000\\
	9.30000000000000	3.26000000000000e-05\\
	10.3000000000000	3.40000000000000e-06\\
};
\label{lab:5G_20}

\addplot [color=red,line width = 0.5pt,solid,mark=o]
table[row sep=crcr]{
	1	0.937100000000000\\
	2	0.737000000000000\\
	3	0.416600000000000\\
	4	0.155400000000000\\
	5	0.0412000000000000\\
	6	0.00712222222222222\\
	7	0.000879500000000000\\
	8	9.90000000000000e-05\\
	9	1.01800000000000e-05\\
	10	1.10800000000000e-06\\
};
\label{plot:Ray_GenAlg_8dB_20}

\coordinate (legend) at (axis description cs:0.67,0.04);

\draw[<->,black,line width=0.75pt] (6.95,1e-3) -- (7.6,1e-3);
\node[black,text opacity=1,fill=white, fill opacity=0.7] at (7.6,6.2e-4) {\footnotesize $\mathbf{\unit[0.65]{\mathbf{dB}}}$};

\draw[<->,black,line width=0.75pt] (8,1e-4) -- (8.8,1e-4);
\node[black,text opacity=1,fill=white, fill opacity=0.7] at (8.65,6.2e-5) {\footnotesize $\mathbf{\unit[0.8]{\mathbf{dB}}}$};

\draw[<->,black,line width=0.75pt] (9,1e-5) -- (9.8,1e-5);
\node[black,text opacity=1,fill=white, fill opacity=0.7] at (9.54,6.2e-6) {\footnotesize $\mathbf{\unit[0.8]{\mathbf{dB}}}$};

\end{axis}

\matrix [
draw,
fill=white,
matrix of nodes,
anchor=south east,
font=\footnotesize,
mark options={solid}
] at (legend) {
	& Construction @ design SNR \\
	\ref{plot:CCSDS_20} &  CCSDS Up-Link LDPC \cite{NorbertWehn} \\
	\ref{lab:36_20} &  (3,6) Regular LDPC \\
	\ref{lab:5G_20} &  5G LDPC \cite{polar5G2018} \\
	\ref{plot:Ray_GenAlg_8dB_20} &  GenAlg LDPC @ 8 dB \\			
};

\end{tikzpicture}
		\vspace{-0.7cm}
		\caption{BLER comparison}
				\label{fig:Ray_BP20}
	\end{subfigure}
	
	\vspace{0.1cm}
	
	\begin{subfigure}{\columnwidth}
		\begin{tikzpicture}
\begin{axis}[
     width=9cm,
     height=6.7cm,
     grid=both,
     grid style={dotted,anthrazit},
       legend cell align=left,
       legend style={font=\footnotesize},
       legend columns=1,
       legend style={at={(0.001,0.3)},anchor=west,draw=none,
       	/tikz/every even column/.append style={column sep=0.5cm}
       },
	xmin=1,
	xmax=10,
	ymin=1,
	ymax=20,
	ytick={1,5,10,15,20},	
	major tick length=0pt,
	xlabel={$E_b/N_0$ [dB]},
	ylabel={Average Iterations $N_{it,avg}$},
	legend image post style={mark indices={}},
	mark options={solid}
	]

\addplot[color=black,line width = 0.5pt,solid,mark=x]
table[row sep=crcr]{
1	                19.9838000000000\\
1.90000000000000	19.7349000000000\\
2.90000000000000	18.2760000000000\\
3.90000000000000	13.8720000000000\\
4.90000000000000	8.92500000000000\\
5.90000000000000	5.27590000000000\\
6.90000000000000	3.34855555555556\\
7.90000000000000	2.40052650000000\\
8.90000000000000	1.89118830000000\\
9.90000000000000	1.56934047000000\\
10.9000000000000	1.34725531800000\\
	};
	\label{lab:Ray_protonasa_20} 

\addplot[color=blue,line width = 0.5pt,solid,mark=triangle]
table[row sep=crcr]{
1	                19.8384000000000\\
1.60000000000000	19.3501000000000\\
2.60000000000000	16.8458000000000\\
3.60000000000000	12.2291000000000\\
4.60000000000000	7.75750000000000\\
5.60000000000000	4.74550000000000\\
6.60000000000000	3.17623333333333\\
7.60000000000000	2.35168250000000\\
8.60000000000000	1.87743775000000\\
9.60000000000000	1.56176730000000\\
10.6000000000000	1.34259346250000\\
};
\label{lab:Ray_36_20}

\addplot[color=darkgreen,line width = 0.5pt,solid,mark=otimes,mark size=2.5pt]
table[row sep=crcr]{
1	                19.7489000000000\\
1.30000000000000	19.5129000000000\\
2.30000000000000	17.5934000000000\\
3.30000000000000	13.7226000000000\\
4.30000000000000	9.78080000000000\\
5.30000000000000	6.81760000000000\\
6.30000000000000	5.13868000000000\\
7.30000000000000	4.19310500000000\\
8.30000000000000	3.62368440000000\\
9.30000000000000	3.24366500000000\\
10.3000000000000	2.97126726000000\\
};
\label{lab:Ray_5G_20}

\addplot [color=red,line width = 0.5pt,solid,mark=o]
table[row sep=crcr]{
	1	19.4126000000000\\
	2	17.0856000000000\\
	3	12.6891000000000\\
	4	8.15490000000000\\
	5	5.31320000000000\\
	6	3.74278888888889\\
	7	2.92758400000000\\
	8	2.41665780000000\\
	9	2.04930941000000\\
	10	1.76617619000000\\
};
\label{plot:Ray_GenAlg_8dB_20}

\coordinate (legend) at (axis description cs:1,0.48);
\end{axis}

\matrix [
draw,
fill=white,
matrix of nodes,
anchor=south east,
font=\footnotesize,
mark options={solid}
] at (legend) {
	& Construction @ design SNR \\
	\ref{lab:Ray_protonasa_20} &  CCSDS Up-Link LDPC \cite{NorbertWehn} \\
	\ref{lab:Ray_36_20} &  (3,6) Regular LDPC \\
	\ref{lab:Ray_5G_20} &  5G LDPC \cite{polar5G2018} \\	
	\ref{plot:Ray_GenAlg_8dB_20} &  GenAlg LDPC @ 8 dB \\	
};

\end{tikzpicture}
		\vspace{-0.7cm}
		\caption{Average number of required iterations}
				\label{fig:Ray_64128LDPC_avgIter20}
	\end{subfigure}
	
	\vspace{0.3cm}
	
	\begin{subfigure}{\columnwidth}
		\begin{tikzpicture}
\begin{axis}[
     width=9cm,
     height=6.7cm,
     grid=both,
     grid style={dotted,anthrazit},
       legend cell align=left,
       legend style={font=\footnotesize},
       legend columns=1,
       legend style={at={(0.001,0.3)},anchor=west,draw=none,
       	/tikz/every even column/.append style={column sep=0.5cm}
       },
	xmin=1,
	xmax=10,
	ymin=0,
	ymax=160,
	major tick length=0pt,
	xlabel={$E_b/N_0$ [dB]},
	ylabel={Average decoding complexity $\eta$},
	legend image post style={mark indices={}},
	mark options={solid}
	]

\addplot[color=black,line width = 0.5pt,solid,mark=x]
table[y expr=\thisrowno{1}*512/64, x expr=\thisrowno{0}, row sep=crcr]{
1	                19.9838000000000\\
1.90000000000000	19.7349000000000\\
2.90000000000000	18.2760000000000\\
3.90000000000000	13.8720000000000\\
4.90000000000000	8.92500000000000\\
5.90000000000000	5.27590000000000\\
6.90000000000000	3.34855555555556\\
7.90000000000000	2.40052650000000\\
8.90000000000000	1.89118830000000\\
9.90000000000000	1.56934047000000\\
10.9000000000000	1.34725531800000\\
	};
	\label{lab:Ray_protonasa_20} 

\addplot[color=blue,line width = 0.5pt,solid,mark=triangle]
table[y expr=\thisrowno{1}*384/64, x expr=\thisrowno{0}, row sep=crcr]{
1	                19.8384000000000\\
1.60000000000000	19.3501000000000\\
2.60000000000000	16.8458000000000\\
3.60000000000000	12.2291000000000\\
4.60000000000000	7.75750000000000\\
5.60000000000000	4.74550000000000\\
6.60000000000000	3.17623333333333\\
7.60000000000000	2.35168250000000\\
8.60000000000000	1.87743775000000\\
9.60000000000000	1.56176730000000\\
10.6000000000000	1.34259346250000\\
};
\label{lab:Ray_36_20}

\addplot[color=darkgreen,line width = 0.5pt,solid,mark=otimes,mark size=2.5pt]
table[y expr=\thisrowno{1}*473/64, x expr=\thisrowno{0}, row sep=crcr]{
1	                19.7489000000000\\
1.30000000000000	19.5129000000000\\
2.30000000000000	17.5934000000000\\
3.30000000000000	13.7226000000000\\
4.30000000000000	9.78080000000000\\
5.30000000000000	6.81760000000000\\
6.30000000000000	5.13868000000000\\
7.30000000000000	4.19310500000000\\
8.30000000000000	3.62368440000000\\
9.30000000000000	3.24366500000000\\
10.3000000000000	2.97126726000000\\
};
\label{lab:Ray_5G_20}

\addplot [color=red,line width = 0.5pt,solid,mark=o]
table[y expr=\thisrowno{1}*398/64, x expr=\thisrowno{0}, row sep=crcr]{
	1	19.4126000000000\\
	2	17.0856000000000\\
	3	12.6891000000000\\
	4	8.15490000000000\\
	5	5.31320000000000\\
	6	3.74278888888889\\
	7	2.92758400000000\\
	8	2.41665780000000\\
	9	2.04930941000000\\
	10	1.76617619000000\\
};
\label{plot:Ray_GenAlg_8dB_20}

\coordinate (legend) at (axis description cs:1,0.48);

\node[black,text opacity=1, draw, fill=white, fill opacity=1] at (2.25,50) {\small $\eta = \frac{N_{it,avg} \cdot E}{k}$};

\end{axis}

\matrix [
draw,
fill=white,
matrix of nodes,
anchor=south east,
font=\footnotesize,
mark options={solid}
] at (legend) {
	& Construction @ design SNR \\
	\ref{lab:Ray_protonasa_20} &  CCSDS Up-Link LDPC \cite{NorbertWehn} \\
	\ref{lab:Ray_36_20} &  (3,6) Regular LDPC \\
	\ref{lab:Ray_5G_20} &  5G LDPC \cite{polar5G2018} \\	
	\ref{plot:Ray_GenAlg_8dB_20} &  GenAlg LDPC @ 8 dB \\	
};

\end{tikzpicture}
		\vspace{-0.7cm}
		\caption{Average decoding complexity; $E$ is the number of edges in the graph of the code}
				\label{fig:Ray_64128LDPC_avgComp20}
	\end{subfigure}
	\vspace{-0.1cm}	
	\caption{\footnotesize Several $\left(n=128,k=64\right)$ LDPC codes decoded with BP decoding using \underline{$N_{it,max}=20$ iterations} over the \underline{Rayleigh fading channel}.}
		\label{fig:Ray_20_ALL}
\end{figure}

\section{Lessons learned from the genetic learning algorithm}\label{sec:lessons}

Besides superior decoding performance, we aim to \emph{understand} what makes the resulting codes so powerful and further analyze the final code structure. As such, the GenAlg may even provide new design paradigms for short length codes.

First, we observed that our optimized GenAlg-based LDPC codes (i.e., $\mathbf{H}$-matrices) contain some degree-1 VNs. 
It is well-known that degree-1 VNs are unfavorable in unstructured LDPC codes as they increase the probability of having more than one degree-1 VN being connected to the same CN (recall TB-IRA and PTB-IRA).
In that case, these degree-1 VNs suffer from unrecoverable poor decoding performance.
Obviously, this, prevents having an open tunnel between the VND and CND EXIT curves as shown in Fig.~\ref{fig:GenAlg_EXIT}, and converging to the $\left(1,1\right)$ point is not possible. However, as shown in Fig.~\ref{fig:shortlength}, if wisely placed, those degree-1 VNs are not degrading the actual error-rate performance of the short-length LDPC code.
Similarly in \cite{divsalar}, degree-1 VNs were used to improve the iterative decoding thresholds of protograph-based LDPC codes. 
One interesting observation is that in our optimized GenAlg-based LDPC codes, there is no CN which is connected to more than one degree-1 VN, despite the fact that this was not a constraint in the optimization problem. Thus, GenAlg was able to ``learn'' that the design in which a CN is connected to more than one degree-1 VN should be avoided.
A similar application of degree-1 VNs was reported by Richardson in multi-edge type LDPC codes \cite{multi-edge-type-LDPC}. 

In order to further investigate the found degree profile, we show a BER comparison between two long length LDPC codes $\left(n=128000\right)$ in Fig.~\ref{fig:BER_Matched_GenAlg}, as a sanity check:
\begin{enumerate}
	\item \emph{Conventional design:} the first code is a single realization following the (asymptotically) optimal degree profile found by EXIT chart-based curve matching as shown in Fig.~\ref{fig:EXIT_matched}.  
	\item \emph{GenAlg-based design:} the second code is a \emph{scaled} version (realization) following the degree profile of our optimized GenAlg-based short length LDPC code as shown in Fig.~\ref{fig:GenAlg_EXIT}. It has a non-negligible amount of degree-1 VNs, but we ensure per-design that only one degree-1 VN is connected per CN.
\end{enumerate}
As shown in Fig.~\ref{fig:BER_Matched_GenAlg}, it can be seen that the long LDPC code designed based on EXIT chart curve matching benefits from a lower threshold than the code based on GenAlg. Further, the GenAlg-based long LDPC code indeed suffers from an inevitable error-floor due to the significant portion of degree-1 VNs it has.
Therefore, Fig.~\ref{fig:shortlength} and Fig.~\ref{fig:BER_Matched_GenAlg} clearly indicate that short length codes follow \emph{different} design paradigms and, thus, the GenAlg-based design process leads to better (in terms of error-rate) short length LDPC codes. The intuition behind this effect is the fact that, typically, the price-to-pay for non-matched EXIT curves is a degraded waterfall performance. However, in the short length regime, the slope in the waterfall region is more important than its exact starting position (i.e., threshold).

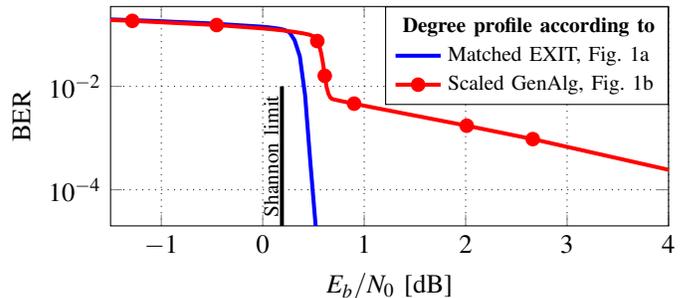
\begin{figure}[t]
	\begin{tikzpicture}
\begin{semilogyaxis}[
     width=9cm,
     height=4.5cm,
     grid=both,
     grid style={dotted,anthrazit},
       legend cell align=left,
       legend style={font=\footnotesize},
       legend columns=1,
       legend style={at={(0.492,0.776)},anchor=west,draw,
       	/tikz/every even column/.append style={column sep=0.5cm}
       },
	xmin=-1.5,
	xmax=4,
	ymin=2e-5,
	ymax=3.5e-1,
	xlabel={$E_b/N_0$ [dB]},
	ylabel={BER},
	legend image post style={mark indices={}},
	mark options={solid}
	]

\addlegendimage{empty legend}
\addlegendentry{\hspace{-.6cm}\textbf{Degree profile according to}}

\addplot[color=blue,line width = 1.5pt,solid]
table[row sep=crcr]{
1.01029698737655	3.22515625e-06\\
0.95469292709908	3.5782833530326e-06\\
0.899452195302141	3.94433206933207e-06\\
0.844550992923911	4.3908872033872e-06\\
0.789994627985099	4.98862998862999e-06\\
0.735778795625266	5.76731826731827e-06\\
0.681899271095314	6.71215514965515e-06\\
0.628351907781952	8.126555001555e-06\\
0.575132635292703	1.03435181560182e-05\\
0.522246681750342	1.92294453125e-05\\
0.46967161972337	0.000328042429749755\\
0.417412877047694	0.00672307043083738\\
0.365466670187034	0.0381334194032486\\
0.313829283084781	0.0781477254231771\\
0.262497065568813	0.111872273763021\\
0.211475331758976	0.123921223958333\\
0.160742706909273	0.130375793457031\\
0.110304681777752	0.134560994466146\\
0.0601578546808699	0.138381876627604\\
0.0102988825148742	0.141460327148438\\
-0.00969672411294686	0.142703450520833\\
-0.553294274946404	0.166300048828125\\
-1.06485812357033	0.184010986328125\\
-1.54796137246529	0.198656677246094\\
-2.00560460098709	0.212078348795573\\
-2.44034409136211	0.224032023111979\\
-2.85435132302107	0.235192016601563\\
-3.24951970881791	0.245291158040365\\
-3.62749508174737	0.254676005045573\\
-3.98970060156	    0.263661193847656\\
};
\label{lab:BER_matched}
\addlegendentry{Matched EXIT, Fig.~\ref{fig:EXIT_matched}};

\addplot [color=red,line width = 1.5pt,solid,mark=*,mark indices={1,2,13,23,27,40,41}]
table[row sep=crcr]{
	4.01029419682456	0.000240672354310203\\
	2.66153504678388	0.000954781668526786\\
	2.01029839824869	0.00173372625073357\\
	1.89272942532746	0.00192764727274577\\
	1.77673059312532	0.00213174985323338\\
	1.66226051510708	0.00235657051282051\\
	1.54927941980499	0.00260746689383865\\
	1.49433587308967	0.00273466435185185\\
	1.43774906786151	0.00287063764232074\\
	1.32763267433103	0.00316327582465278\\
	1.21889483584532	0.0034839654369889\\
	1.11150146228198	0.00381888037612758\\
	1.00541971260591	0.00420701479108146\\
	0.900617934581545	0.00462603985821759\\
	0.797065608079112	0.00508097220755912\\
	0.760298535590305	0.00525767856174045\\
	0.741519106828125	0.00536671691009964\\
	0.722789632368037	0.00545986186594203\\
	0.704091038276036	0.0055644677005597\\
	0.685432611225386	0.00568019982540246\\
	0.666814179017745	0.00604593387726815\\
	0.648244929492326	0.00713943884507665\\
	0.629705954831908	0.00961556615584936\\
	0.611206465096541	0.0159933420817057\\
	0.592755591785171	0.0289206368582589\\
	0.574334549760964	0.0452909749348958\\
	0.555952492433922	0.0614523010253906\\
	0.537609255144788	0.0752530314127604\\
	0.519313895312441	0.0871735432942708\\
	0.501047788905488	0.0924380900065104\\
	0.482820014850976	0.0957920939127604\\
	0.46463041260099	0.0991128946940104\\
	0.446487966661706	0.101315999348958\\
	0.428374211345404	0.103008280436198\\
	0.410298152274754	0.104871032714844\\
	0.394777135859876	0.106210530598958\\
	0.297050322041914	0.113990173339844\\
	0.200410831277309	0.120424560546875\\
	0.10483473412866	0.126175455729167\\
	0.0102988825148742	0.131502095540365\\
	-0.45416649150171	0.153409118652344\\
	-1.28575964162589	0.184856160481771\\
	-2.04464583589494	0.209419230143229\\
	-2.74252283815581	0.229845540364583\\
	-3.3884802158291	0.247248128255208\\
	-3.98970060156	0.262696899414062\\
};
\addlegendentry{Scaled GenAlg, Fig.~\ref{fig:GenAlg_EXIT}}

\addplot[color=black,line width = 1.5pt,solid]
table[row sep=crcr]{%
	0.189 1e-2\\ 
	0.189 1e-8\\
};

\node [rotate=90] at (axis cs:0.085,4.5e-4) {\footnotesize Shannon limit};    

\end{semilogyaxis}
\end{tikzpicture}
	\vspace{-0.8cm}
	\caption{\footnotesize Sanity check for long codes; $\left(n=128000,k=64000\right)$ LDPC codes; BP decoding with $N_{it,max}=200$; bi-AWGN channel.} 
	\label{fig:BER_Matched_GenAlg}	
	\vspace{-0.4cm}
\end{figure}

Further, we also observe that in our GenAlg-based short LDPC codes, no degree-2 VNs are involved in a short cycle. Despite being a well-known structural constraint in LDPC design, this was not explicitly outlined as a constraint in our GenAlg optimization process. Again, GenAlg was able to ``\emph{learn}'' this constraint independently.

Next we summarize a few remarkable observations from our GenAlg results:
\begin{itemize}
	\item Degree-1 VNs do not necessarily cause a performance degradation in short length LDPC codes.
	\item In the short length LDPC code design process, a non-matched EXIT curves scenario can lead to a good short length LDPC code, while obviously leading to a poor long length LDPC code.
	\item Unfavorable graph structures (e.g., degree-2 VNs involved in short girth) were inherently avoided by GenAlg without an explicit constraint imposed on it.
\end{itemize}  

\section{Conclusion}

The classical LDPC code design tools are based on asymptotic length assumptions which are not valid in the short-length regime. 
Therefore, we focus on constructing short-length LDPC codes (i.e., the parity-check matrix) using the genetic algorithm.
We propose a flexible framework accommodating practical decoding requirements and channel constraints.
We construct LDPC codes without any special graph structure (i.e., we use a random edge interleaver) and demonstrate the flexibility of the proposed framework. We also construct accumulator-based LDPC codes which can be encoded easily.
Our proposed LDPC codes outperform some well-designed state-of-the-art (standardized) LDPC codes over both AWGN and Rayleigh fading channels.
Moreover, we design LDPC codes tailored to a reduced number of BP iterations in order to reduce the decoding complexity and latency with good error-rate performance (e.g., a coding gain of up to $\unit[0.8]{dB}$ when compared to the 5G LDPC code is reported over the Rayleigh fading channel). Finally, we observed that allowing the presence of carefully placed degree-1 VNs opens up more degrees of freedom for code design, and does not degrade the error-rate performance of our proposed short length LDPC codes.

\pagebreak

\bibliographystyle{IEEEtran}
\bibliography{references}

\begin{IEEEbiography}[{\includegraphics[width=1in,height=1.25in,clip,keepaspectratio]{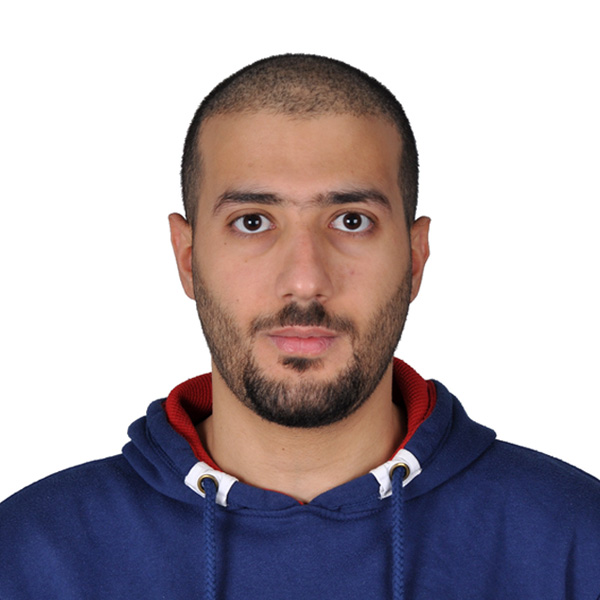}}]
	{Ahmed Elkelesh} received the B.Sc. degree (with highest honours) in Information Engineering and Technology in 2013 from the German University in Cairo and M.Sc. degree (with distinction) in Communications Engineering and Media Technology in 2016 from the University of Stuttgart. 
	During his years of study in Germany, he was a research assistant with Fraunhofer IPA Stuttgart and an intern at Sony Stuttgart Technology Center.
	Since 2016, he has been a member of research staff with the Institute of Telecommunications, University of Stuttgart, where he is working toward the Ph.D. degree. 
	His main research topic is channel coding, with particular emphasis on polar codes and LDPC codes.
	Further research interests include the areas of information theory, modulation, machine learning and SDR.
	He was the recipient of the Anton-und-Klara-R{\"o}ser prize 2017 for his master thesis. 
\end{IEEEbiography}

\vspace{-1.1cm}

\begin{IEEEbiography}[{\includegraphics[width=1in,height=1.25in,clip,keepaspectratio]{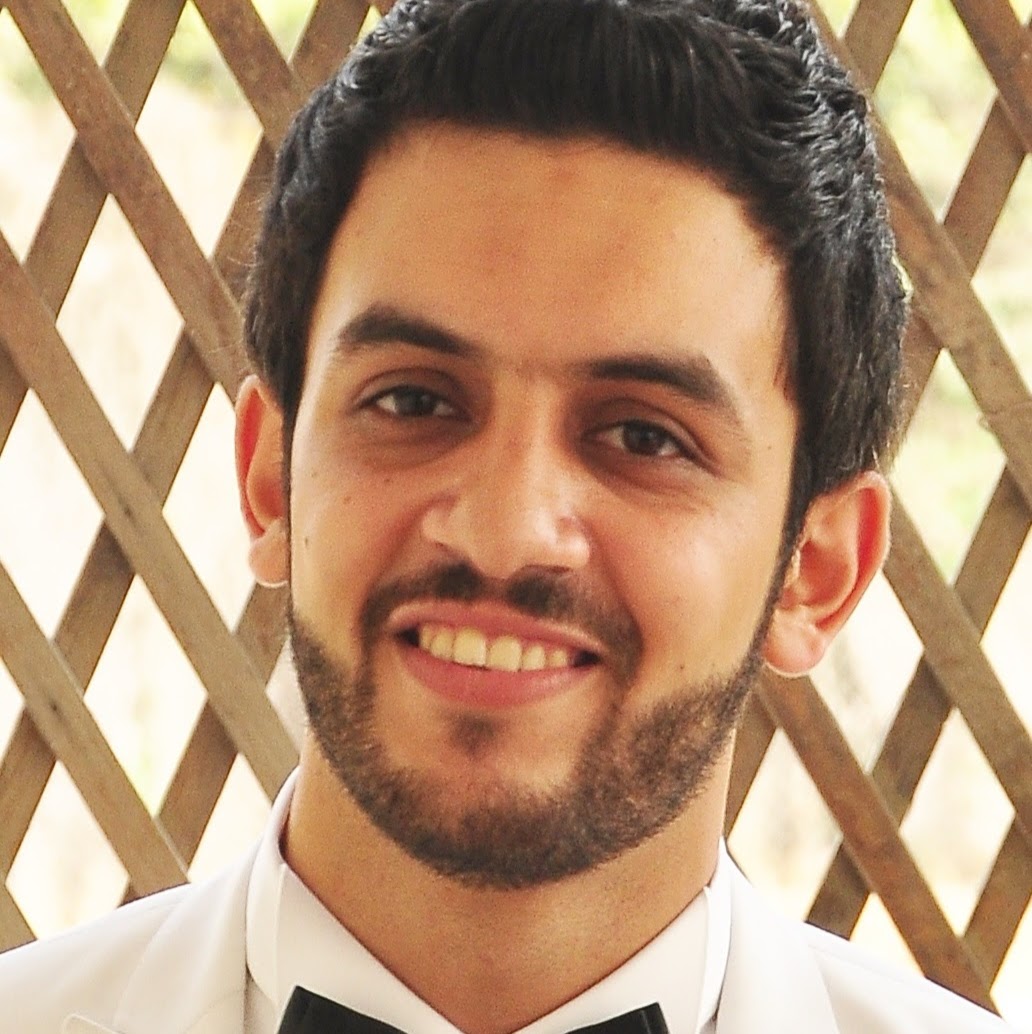}}]
	{Moustafa Ebada} received the B.Sc. (with distinction) from the Communications department in the German University in Cairo, Egypt in 2013 and his M.Sc. degree in electrical engineering and information technology from the University of Stuttgart, Germany in 2016, where he is currently working toward the Ph.D. degree. During his master studies, he was a research assistant with multiple institutes of the University of Stuttgart including fields of radio frequency technology, signal processing and telecommunications. Since 2016, he has been a member of research staff with the Institute of Telecommunications, University of Stuttgart. His main research topics are channel coding, particularly polar code construction and decoding. Besides, designing short LDPC codes for high speed applications. Further research interests include machine learning, particularly designing error correction codes and utilization of the state-of-the-art coding schemes in the field of Multiple Access Channel.  
\end{IEEEbiography}

\vspace{-1.1cm}

\begin{IEEEbiography}[{\includegraphics[width=1in,height=1.25in,clip,keepaspectratio]{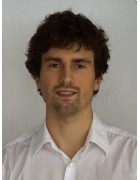}}]
	{Sebastian Cammerer} (S'16) received the B.Sc. and M.Sc. degree (with 
	distinction) in electrical engineering and information technology from 
	University of Stuttgart, Germany, in 2013 and 2015, respectively.
	During his years of study he worked as a research assistant at multiple 
	institutes of University of Stuttgart.
	Since 2015 he is a member of research staff at Institute of 
	Telecommunications, University of Stuttgart, and is pursuing his Ph.D.
	His main research topics are channel coding and machine learning for 
	communications.
	Further research interests are in the areas of modulation, parallelized 
	computing for signal processing and information theory.
	He is recipient of the Best Publication Award of the University of Stuttgart 2019, the Anton- und Klara R{\"o}ser Preis 2016, the Rohde\&Schwarz Best Bachelor Award 2015 and the VDE-Preis 2016 for his master thesis.
\end{IEEEbiography}

\vspace{-1.1cm}

\begin{IEEEbiography}[{\includegraphics[width=1in,height=1.25in,clip,keepaspectratio]{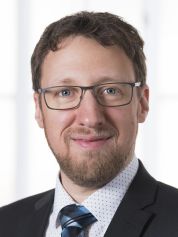}}]
	{Laurent Schmalen} is a professor at Karlsruhe Institute of Technology 
	(KIT) where he leads the Communications Engineering Lab (CEL). He 
	received the Dipl.-Ing. degree in electrical engineering and information 
	technology and the Dr.-Ing. degree from RWTH Aachen University of 
	Technology, Aachen, Germany. From 2011 to 2019, he was with Bell Labs as 
	a member of technical staff and from 2016 to 2019 also as department 
	head of the Coding in Optical Communications department. From 2014-2019, 
	he was Guest Lecturer at the University of Stuttgart, Germany. His 
	research interests include forward error correction, modulation formats, 
	and information theory for future optical networks. He is a senior 
	member of the IEEE and recipient and co-recipient of several awards, 
	including the E-Plus Award for his Ph.D. thesis, the Best Paper 
	Award of the 2010 ITG Speech Communication Conference, the 2013 Best 
	Student Paper Award at the IEEE Signal Processing Systems workshop, and 
	the 2016 Journal of Lightwave Technology best paper award. Additionally, 
	he received 2014 IEEE Transactions on Communications Exemplary Reviewer 
	Award and the 2019 Journal of Lightwave Technology Outstanding Reviewer 
	Recognition.
\end{IEEEbiography}

\begin{IEEEbiography}[{\includegraphics[width=1in,height=1.25in,clip,keepaspectratio]{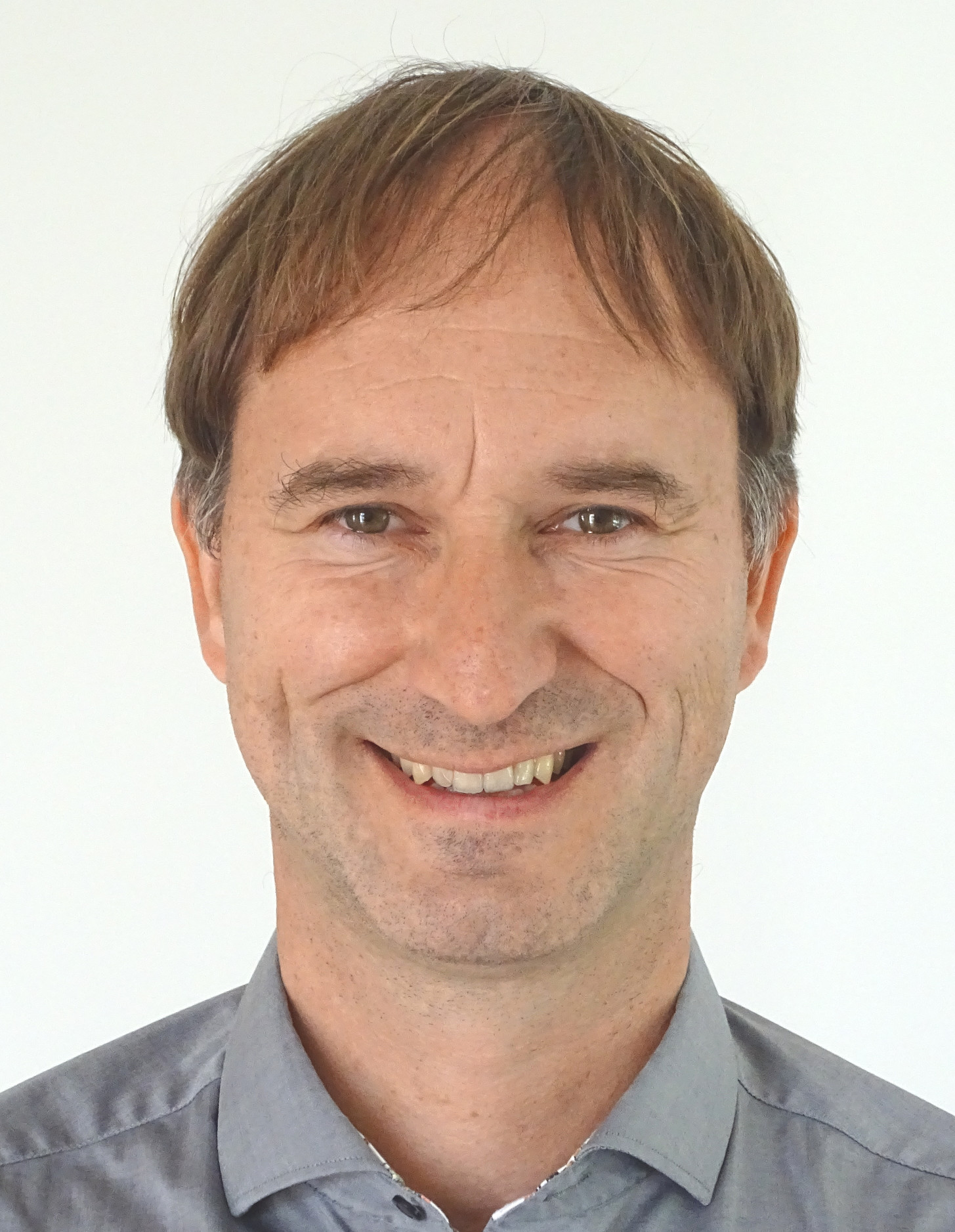}}]
	{Stephan ten Brink}(M'97--SM'11) has been a faculty member at the 
	University of Stuttgart, Germany, since July 2013, where he is head of 
	the Institute of Telecommunications.
	From 1995 to 1997 and 2000 to 2003, Dr. ten Brink was with Bell 
	Laboratories in Holmdel, New Jersey, conducting research on multiple 
	antenna systems.
	From July 2003 to March 2010, he was with Realtek Semiconductor Corp., 
	Irvine, California, as Director of the wireless ASIC department, 
	developing WLAN and UWB single chip MAC/PHY CMOS solutions.
	In April 2010 he returned to Bell Laboratories as Department Head of the 
	Wireless Physical Layer Research Department in Stuttgart, Germany.
	Dr. ten Brink is a recipient and co-recipient of several awards, 
    including the Vodafone Innovation Award, the IEEE Stephen O. Rice Paper Prize, 
    the IEEE Communications Society Leonard G. Abraham Prize for contributions to channel coding and 
	signal detection for multiple-antenna systems.
	He is best known for his work on iterative decoding (EXIT charts) and 
	MIMO communications (soft sphere detection, massive MIMO).
\end{IEEEbiography}

\end{NoHyper}
\end{document}